\documentclass[manuscript]{acmart}

\usepackage{subcaption}
\usepackage{multirow}
\usepackage{float}

\AtBeginDocument{%
  }

\setcopyright{acmlicensed}
\copyrightyear{2025}
\acmYear{2025}
\begin{document}

\title{Wild Narratives: Exploring the Effects of Animal Chatbots on Empathy and Positive Attitudes toward Animals}

\author{Jingshu Li}
\authornote{All authors contributed equally to this research.}
\email{jingshu@u.nus.edu}
\affiliation{%
  \institution{Computer Science, National University of Singapore}
  \city{Singapore}
  \country{Singapore}
}

\author{Aaditya Patwari}
\authornotemark[1]
\affiliation{%
  \institution{Yale-NUS}
  \city{Singapore}
  \country{Singapore}}
\email{aadipatwari@gmail.com}

\author{Yi-Chieh Lee}
\authornotemark[1]
\email{yclee@nus.edu.sg}
\affiliation{%
  \institution{Computer Science, National University of Singapore}
  \city{Singapore}
  \country{Singapore}
}


\begin{abstract}

Rises in the number of animal abuse cases are reported around the world. While chatbots have been effective in influencing their users’ perceptions and behaviors, little if any research has hitherto explored the design of chatbots that embody animal identities for the purpose of eliciting empathy toward animals. We therefore conducted a mixed-methods experiment to investigate how specific design cues in such chatbots can shape their users’ perceptions of both the chatbots’ identities and the type of animal they represent. Our findings indicate that such chatbots can significantly increase empathy, improve attitudes, and promote prosocial behavioral intentions toward animals, particularly when they incorporate emotional verbal expressions and authentic details of such animals’ lives. These results expand our understanding of chatbots with non-human identities and highlight their potential for use in conservation initiatives, suggesting a promising avenue whereby technology could foster a more informed and empathetic society.

\end{abstract}





\maketitle

\section{Introduction} 

Many countries have recently reported rises in animal abuse \cite{Maisner2023, Cheng2024, Harris2023}. Various factors have been identified as contributing to this troubling trend, including a lack of awareness that animals can suffer pain \cite{Agnew1998}. Another factor is absence of empathy for animals, as individuals who do not recognize them as sentient are less likely to treat them humanely~\cite{Miller2001, Agnew1998}. In addition, the impossibility of communicating with animals verbally leaves them particularly susceptible to misrepresentation and misunderstanding, further increasing their vulnerability to abuse~\cite{Thomsen2023}.

Researchers have explored various approaches to promoting people’s empathy and positive attitudes toward animals. These range from live animal encounters in educational settings like zoos~\cite{Malecki2019} to first-person narratives purportedly written by animals \cite{Akerman2019}. A recurring theme across these strategies is the attempt to provide animals with a voice, as a means of evoking empathy and thus helping to shift public perception in favor of better treatment of animals \cite{Young2018, Demello2018, Malecki2019}. Additionally, some efforts to improve human empathy with animals have involved showing animals using digital tools such as iPads \cite{webber2017interactive}. The results suggest that seeing animals as human-like indeed fosters empathy by heightening people’s awareness that animals also have thoughts and feelings~\cite{webber2017kinecting}. However, some research has raised concerns that the use of technology in animals is unnatural or increases animal stress \cite{diana2021systematic, webber2017kinecting}.

Extensive research has focused on the ability of conversational agents (also called chatbots) to shape people’s perceptions and engagement through a variety of narrative techniques~\cite{Bickmore2009, Park2023}. To enhance chatbots’ facilitation of prosocial attitudinal and behavioral changes~\cite{Park2023}, as well as their ability to provide mental-health support~\cite{Bickmore2009, Gennaro2020, lee2019caring}, scholars have explored methods of humanizing them: for instance, assigning them names and/or portraits, and designing them to exhibit emotions~\cite{Shi2020, Liu2018, Seeger2018}. Humanizing chatbots may amplify the positive effects of other aspects of their design. For example, previous research has indicated that chatbots’ employment of first-person narratives can, in addition to increasing empathy and fostering attitudinal changes toward the characters represented in such narratives~\cite{Bickmore2009}, increase user enjoyment during interactions~\cite{lee2023exploring}.

In short, chatbots have demonstrated their ability to influence users’ perceptions across various domains \cite{Park2023, Gennaro2020, lee2019caring}, and research to date suggests that a narrative from an animal’s first-person viewpoint could be effective in promoting empathy and changing human attitudes toward the animal, or toward animals in general \cite{Beierl2008, James2019, Keen2006}. Additionally, we suggest that chatbots could serve as a useful means of simulating human-animal interactions without jeopardizing animal welfare. Nevertheless, research on whether and how chatbots could be designed to represent animals, and how such representation might influence users’ empathy and attitudes toward animals, has hitherto been rare to nonexistent.

To address this absence, we designed chatbots that speak from the perspective of an animal. Inspired by prior work on chatbots’ humanized design cues (e.g., identity, non-verbal and verbal cues) \cite{Seeger2018, Shi2020,Liu2018}, we were interested in whether and how such cues could be redesigned to represent animals, and how this would affect user perceptions of chatbots’ identities. We therefore conducted a mixed-methods experiment with 240 participants, each of whom was randomly assigned to one of eight conditions, i.e., interaction with a chatbot that had zero, one, two, or all of the three above-mentioned cue types. All groups then engaged in interactive dialogue with their respective chatbots. We evaluated the participants’ empathy, attitudes, and intention to engage in prosocial behavior toward the animals portrayed in these narratives. Our results showed that a chatbot’s utilization of verbal cues enhanced participants’ empathy, whereas the use of non-verbal ones resulted in reduced empathy. Notably, first-person narratives by the chatbots succeeded in boosting participants’ perception of the chatbot’s identity as animal-like, and this led to more positive attitudes and increased prosocial behavioral intentions.

This pioneering study makes the following contributions to the human-computer interaction community. First, it extends exploration of human-like chatbots into the domain of animal-like ones, and thus opens new pathways to understanding how non-human personas can be effectively utilized. Second, it demonstrates the ability of chatbots to influence their users’ perceptions of non-human entities. Specifically, we found that chatbots embodying an ‘animal perspective’ could significantly improve user empathy, attitudes, and prosocial behavioral intentions toward animals, particularly when they employed emotional expressions. Finally, this study provides a unique perspective on how human-chatbot interactions can be designed to have positive social impacts. These insights could pave the way for innovative applications in education and conservation, underscoring the potential of chatbot technology to contribute to more empathetic and better-informed societies.


\section{Related Work} 

\subsection{Cruelty to Animals}
Animal abuse can be defined as any action that compromises the well-being of an animal, irrespective of intent, necessity, or social acceptability \cite{Agnew1998}. Previous studies have identified various factors that may contribute to individuals’ cruelty toward animals. One is lack of awareness or comprehension regarding the abusive nature of their actions \cite{Agnew1998}, which may be due to ignorance about how a behavior indirectly contributes to cruelty, or to a belief that animals are incapable of experiencing emotions and/or suffering \cite{Agnew1998}. Another key factor is lack of empathy, including empathy with animals in distress \cite{Agnew1998, Miller2001, McPhedran2009}.

Species that share greater biological similarities with humans, like chimpanzees and bonobos, tend to elicit higher levels of empathy and moral consideration, as evidenced by high levels of public support for their protection \cite{Miralles2019, Agnew1998}. However, this bias leaves more evolutionarily distant animals, such as reptiles and invertebrates, exposed to exploitation in research, the exotic-pet trade, and factory farming \cite{marshall2020thousands}. Further compounding this vulnerability is the communication barrier. The inability to verbally communicate their needs and experiences leaves animals susceptible to misrepresentation and manipulation, as when owners downplay signs of illness or distress in their pets \cite{mellor2016updating}.

\subsection{Representing Animals through First-person Narratives}
A popular method of fostering empathy towards animals is the use of animal narratives \cite{Young2018, Malecki2019}. Framed in ways that resonate with human experiences and emotions, these stories are intended to bridge the gap in understanding between species~\cite{Demello2018}. As DeMello~\cite{Demello2018} has argued, animal narratives can make the internal lives of animals – their desires, fears, and joys – vivid and relatable for human readers, ultimately promoting both a sense of connection and moral consideration.

Various works of literature offer first-person animal narratives. These include Anna Sewell’s {\it Black Beauty} and Leo Tolstoy’s {\it Strider}, both of which feature first-person horse narratives of grueling labor and heartbreaking separation, and Franz Kafka’s {\it Investigations of a Dog}, which offers a dog’s perspective on the world. Despite their fictional nature, these narratives can evoke a sense of sadness and indignation in readers, fostering empathy and potentially improving their attitudes toward animals \cite{Beierl2008, James2019}.

This phenomenon can be explained by the concept of narrative empathy, i.e., the empathy experienced by readers toward characters in a book \cite{Keen2006}. It allows readers to step into the shoes of a character, experiencing the world through their senses and emotions. In the case of animal narratives, narrative empathy has been argued to be an important mechanism whereby these narratives can promote empathy toward animals \cite{James2019}. In recognition of this, the term narrative interspecies empathy has been introduced \cite{Malecki2019}. Importantly, empathy evoked toward a single animal character can result in more positive attitudes toward that animal’s entire species, or even animals belonging to other species~\cite{Malecki2019}.

However, animal narratives extend beyond literature to educational settings that offer live animal encounters, such as zoos \cite{Akerman2019, robovsky2020zoos}. Research suggests that when educators in such contexts employ narrative techniques that emphasize animal individuality, this can foster more respectful attitudes among visitors~\cite{Akerman2019}. For example, instead of simply stating facts about a hesitant crab, an educator could personalize its behavior with a relatable narrative: "\textit{We are going to let him be, he does not want to come out of his home today}" \cite{Young2018}. This approach highlights the crab’s agency and potentially reduces its objectification, thus fostering respect \cite{Young2018}.
This approach aligns well with the latest developments in the role of zoos: i.e., that education and fostering positive connections with animals are key aspects of their mission~\cite{robovsky2020zoos}.

\subsection{Representing Animals through Anthropomorphism}
A common theme in narratives aimed at fostering empathy toward animals is the use of \textit{Anthropomorphism}, the ascription of human-like traits to non-human entities~\cite{Waytz2010}. As humans cannot access the cognitive states of animals, the creation of an animal narrative requires the authors to project human-like abilities and cognition onto their animal characters~\cite{Barcz2015, Demello2018}. For example, \textit{Black Beauty} depicts horses with human-like speech, emotions, and desires \cite{Poirier2019}. Such anthropomorphism has been argued to be a crucial step in fostering pro-animal attitudes by making animals more relatable \cite{Young2018, Demello2018}.

However, excessive anthropomorphism of animals can result in false empathy, i.e., feelings of empathy based on inaccurate perceptions \cite{Young2018, Akerman2019}. For example, when visiting animals in a zoo during winter, some visitors may develop an inaccurate perception that the animals are feeling as cold as they themselves are \cite{Akerman2019}. In some cases, such false empathy can be detrimental to the well-being of humans and animals alike\cite{Young2018}. Therefore, responsible use of anthropomorphism, prioritizing the interests of animals, is crucial \cite{Young2018, Akerman2019}.

Virtual pets, i.e., digital simulations of animals, have evolved beyond their original role as mere substitutes for real pets \cite{Cruces2012}. Research has shown that people who anthropomorphize virtual pets treat them as having life, mental states, and sociality; and in some cases, this could lead to such pets providing an important sense of companionship, and even emotional support and learning opportunities~\cite{melson2009robotic}. Virtual pets have also been found to promote positive behaviors and attitudes, including healthy eating habit~\cite{Byrne2012}, stronger adherence to social norms~\cite{Hildmann2008}, and learning motivation~\cite{Chen2011, o2024design}. Nevertheless, it remains unclear whether these virtual animals can cultivate empathy with animals and/or pro-animal attitudes.

Chatbots have considerable potential to drive positive social change. For example, they have been used to promote prosocial behavior \cite{Park2023, Namkoong2023}, provide companionship~\cite{Skjuve2021, Zhou2020}, and offer mental-health interventions \cite{Bickmore2010, Gennaro2020}. A key consideration when designing chatbots aimed at influencing human behavior is their level of anthropomorphism. This is because chatbots with human-like characteristics tend to foster engagement~\cite{Bickmore2009}, emotional connection \cite{araujo2018living}, and reciprocal behavior~\cite{liang2024dialoging, lee2019caring}. For example, participants in one study~\cite{lee2019caring} offered care and support to a chatbot that expressed its own emotions and revealed its past mistakes, a process that resulted in them developing more compassion for themselves.

Researchers have explored various techniques for making chatbots more human-like. In addition to conferring them with emotions \cite{Liu2018}, these have included giving them human-like names \cite{Shi2020} and pictures \cite{Lee2023}. Seeger et al. \cite{Seeger2018} synthesized prior findings into a framework for anthropomorphic chatbot design, which relies on three types of cues: \textit{verbal}, \textit{nonverbal}, and \textit{identity}. Verbal cues are words and sentences used by the chatbot, generally though not always in textual form. Researchers have explored various strategies to increase human-likeness involving verbal cues, such as the use of social dialogue, emotional expressions, and personal pronouns \cite{Seeger2018}. For example, the expression of compassion has yielded positive results for chatbots designed to provide mental-health support \cite{Gennaro2020, Liu2018}. Nonverbal cues, meanwhile, pertain to communicative expressions and behaviors that do not involve speech or writing. In general, this may encompass a variety of elements like facial expressions and bodily movements \cite{Derrick2011}. However, as text-based chatbots lack physical presence, nonverbal cues are limited to more subtle actions such as lengthened response times or the use of emojis \cite{Seeger2018}. The use of such nonverbal cues has been found lead to chatbots being perceived more positively, e.g., as having higher credibility or greater warmth \cite{Yu2024, Beattie2020}. Lastly, identity cues refer to the information provided by the chatbot that communicates its identity. This could involve the use of human names, profile pictures, and other demographic details like gender and race \cite{Seeger2018}. Prior research has found that declaring a chatbot to have a human identity leads to more human-like perceptions of it, but that this comes at the cost of higher expectations about what it can accomplish \cite{Go2019, Mozafari2020}.

In short, previous research has indicated that presenting information from an animal’s perspective can significantly influence people’s empathy and attitudes toward that animal, its species, or animals in general, and that human-like chatbots can cultivate positive attitudes and behavioral changes toward humans. However, it remains unclear how well a chatbot representing an animal, narrating in the first person, can evoke empathy and foster positive attitudes toward that animal or its species. Filling this gap in our knowledge could pave the way for innovative chatbot designs that enhance people’s understanding of animal conservation and aid in combating animal cruelty.

\subsection{Research Questions}

While prior research has demonstrated the persuasive potential of chatbots that emulate human characteristics, such human-like chatbots are inherently limited in their ability to provide the perspectives of animals, and thus to promote a sense of connection with animals. Therefore, we integrated insights from the existing bodies of research on animal narratives and human-like chatbots to develop an animal-like chatbot that spoke from the perspective of an animal (hereafter referred to as an \textit{animal chatbot}). In designing it, we adapted the same three cues typically used in the development of human-like chatbots, i.e., verbal, non-verbal, and identity cues. We then conducted a mixed-methods experiment to evaluate the effectiveness of this animal chatbot at influencing its users’ empathy with and perceptions of animals. More specifically, we sought answers to the following three research questions (RQs):

{\bf RQ1:} How do the designs of animal chatbot’s verbal, non-verbal, and identity cues affect its users’ perceptions of its identity?

{\bf RQ2:} To what extent does the design of animal chatbot affect its users’ empathy, attitudes, and prosocial behavioral intentions toward the animal it represents?

{\bf RQ3:} To what extent do users’ perceptions of animal-chatbot identity affect their empathy, attitudes, and prosocial behavioral intentions toward the animal represented?

\section{Method}
\subsection{Chatbot Design}
To address our research questions, we developed a chatbot that speaks from the perspective of a fictional horse. Given that animals are perceived differently based on their traits \cite{Agnew1998, Miralles2019}, we considered horses to represent a viable middle ground between animals that are already perceived very positively, such as dogs and cats, and those that are perceived negatively, such as insects \cite{serpell2002pets}. In addition, because horses are often subjected to abuse and cruelty by humans~\cite{Corey2012}, the chatbot could positively influence attitudes toward them and potentially promote better treatment practices.

\subsubsection{Conversation Design}
The animal chatbot provided its users with the first-person perspective of a horse learning to wear a saddle, bridle and stirrups and carry humans for the first time. The narrative drew inspiration from Anna Sewell’s popular novel \textit{"Black Beauty"} \cite{Sewell1877}. The novel’s first-person narrative from the perspective of a horse enables an emotional connection with the audience, a technique that we felt could be mirrored in animal chatbot design. The novel’s underlying messages of empathy, compassion, and moral responsibility~\cite{gavin2017feeling} also resonate with the core principles of ethical chatbot design.

The animal-chatbot conversation we designed consisted of three primary segments. The initial segment involved introductions, during which the chatbot introduced itself as a horse residing on a farm (\textit{“Let me share a bit more about myself. I am a horse, and I live with my owner and many other animals on a farm”}), and allowed the user to introduce themself as well. 
During the second segment, the animal chatbot narrated its experience of learning to wear a saddle, etc., and carry humans: \textit{“Every day, I am supposed to carry my owner around the farm and help him with his work. While doing that, I also have to wear a lot of things, such as a bridle, bit, and saddle”}. The chatbot also commented on its feelings about this: \textit{“I remember the first time I learned how to do all of this. It was such a painful experience”}. Then, in the final segment, the animal chatbot made some final remarks to the user before ending the conversation. For instance: \textit{“I have to go now. Thank you for listening to me and I hope we get to speak again some day. Goodbye!”}

To boost their engagement levels, users were given ample opportunities to interact with the animal chatbot by responding to prompts or asking questions. Where the conversation included yes/no questions, the participants were given predetermined response options. For example, in the initial segment of the conversation, when the animal chatbot posed the question, \textit{“Do you also live and work on a farm?”}, its interlocutor could choose one of the following three response options: \textit{“yes”}, \textit{“no”}, and \textit{“I would rather not say”}. Questions designed to encourage participants to self-reflect and share their opinions, on the other hand, were presented without predefined response options. For example, while discussing the challenging experience of being ridden for the first time, the animal chatbot asked users, \textit{“Can you imagine what it feels like to have a piece of hard steel shoved inside your mouth?”}, and allowed them to respond freely. Users were also provided with several opportunities to ask the animal chatbot open-ended questions, facilitating personalized interactions (please refer to the supplemental materials for an example of a whole dialogue).

\begin{figure}[t]
  \centering
  \begin{subfigure}[t]{0.2\textwidth}
    \centering
    \includegraphics[width=\linewidth]{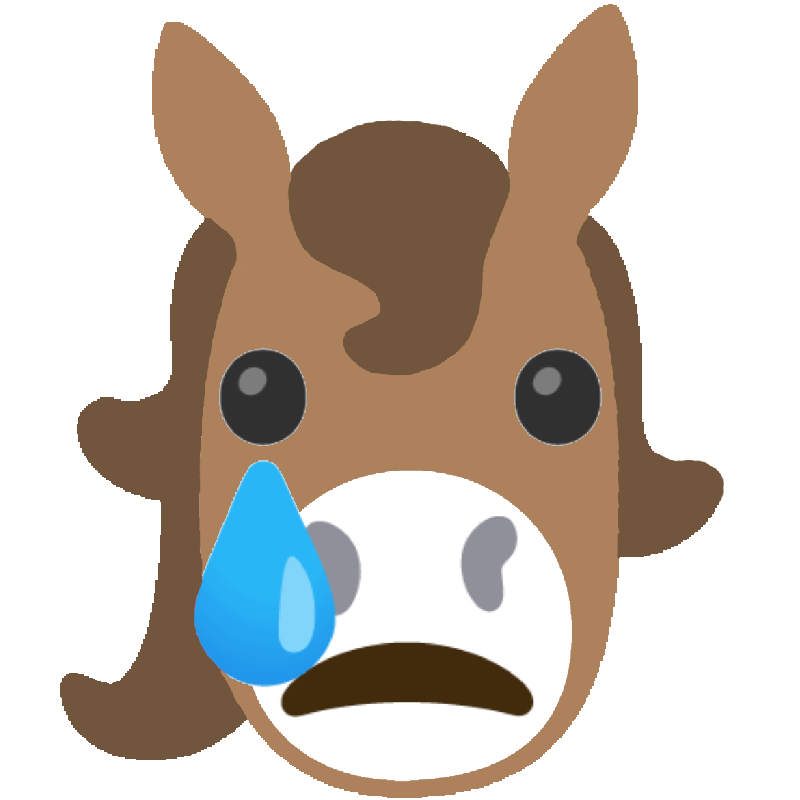}
  \end{subfigure}
  \hfill
  \begin{subfigure}[t]{0.2\textwidth}
    \centering
    \includegraphics[width=\linewidth]{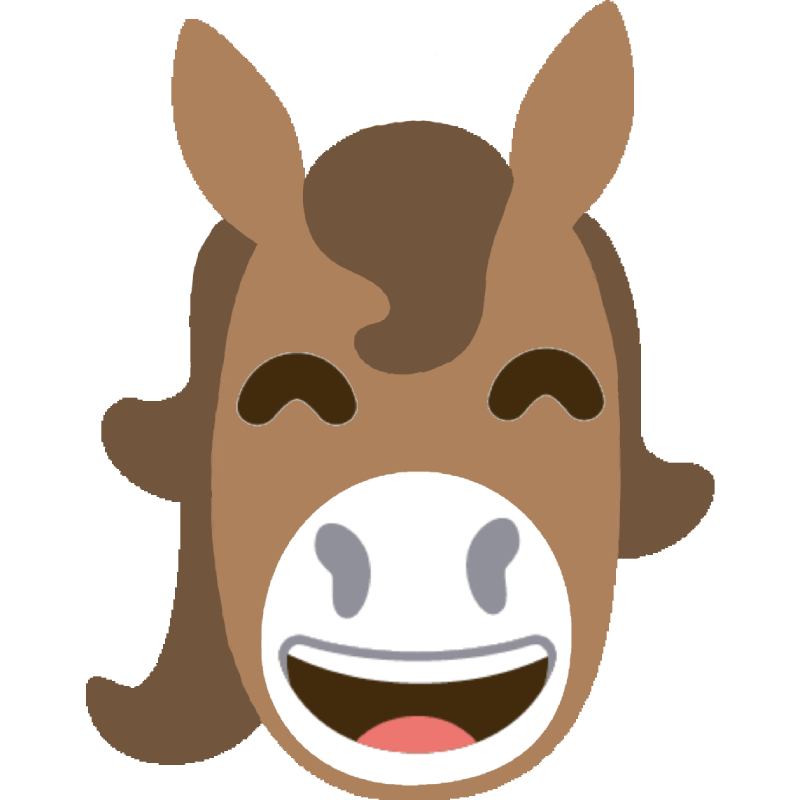}
  \end{subfigure}
  \hfill
  \begin{subfigure}[t]{0.2\textwidth}
    \centering
    \includegraphics[width=\linewidth]{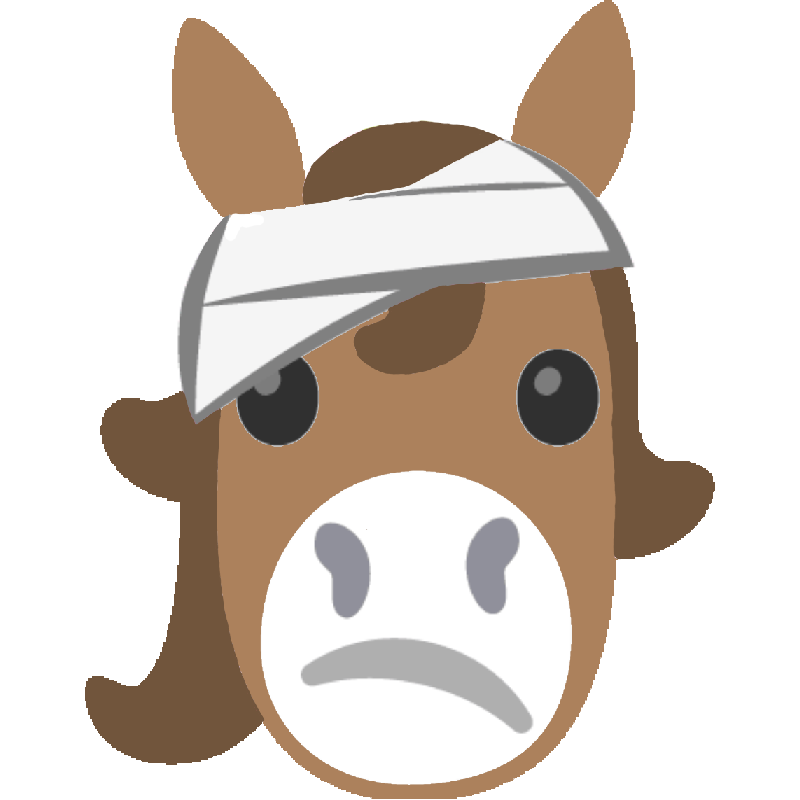}
  \end{subfigure}
  \hfill
  \begin{subfigure}[t]{0.2\textwidth}
    \centering
    \includegraphics[width=\linewidth]{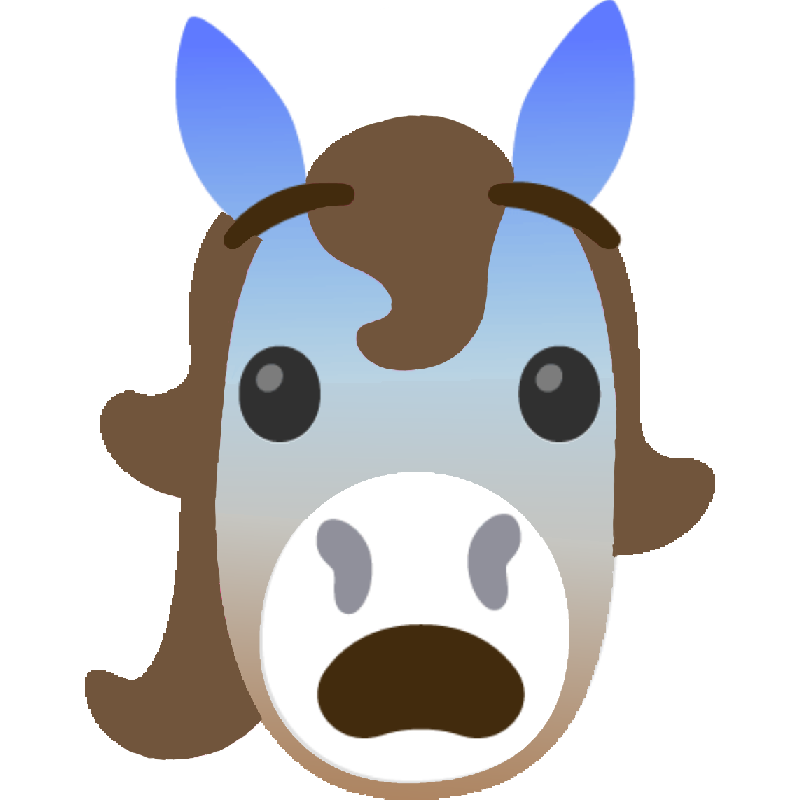}
  \end{subfigure}
  \caption{Example of emojis used as nonverbal cues}
  \label{fig:emojis}
  \Description{This figure shows four horse head emojis, used to convey different emotional states as nonverbal cues. The first emoji depicts a horse with a tear on its cheek, indicating sadness or crying. The second emoji shows a smiling horse with an open mouth, representing happiness or laughter. The third emoji features a horse with a bandage on its head and a sad expression, suggesting injury or pain. 
  The fourth emoji shows a horse with a blue face, suggesting surprised or scared.}
\end{figure}

\subsubsection{Integrating Verbal, Nonverbal, and Identity Cues}
To facilitate our exploration of the impact of verbal, nonverbal, and identity cues on perceptions of animals~\cite{Seeger2018}, we integrated three cues into our animal chatbot and allowed each to be dynamically enabled or disabled (Fig.~\ref{fig:chatbot-interface}). We tailored each cue to be more specific to animal rather than human attributes, as explained below.

{\bf Verbal Cues.}
Inspired by prior literature, we implemented verbal cues in our animal chatbot through the expression of emotions \cite{Seeger2018, bernaerts2014storied}. The various human-like emotions previously incorporated into chatbots have included gratitude~\cite{Namkoong2023} and empathy~\cite{Liu2018}. Similarly, human-like emotions have been used in animal narratives to increase empathy and improve attitudes toward animals \cite{Demello2018, Young2018, Malecki2019, Sewell1877}. In our case, these expressions encompassed emotions ranging from excitement (e.g., \textit{“I am really excited to share more about my life!”}) to discomfort (e.g., \textit{“the shoes felt very heavy”}) and even pain (e.g., \textit{“I felt a sharp, searing pain on my back”}). Our chatbot versions that lacked verbal cues did not utilize any emotional expressions.

More specifically, when describing the experience of learning to wear a bit, the animal chatbot with verbal cues enabled stated, \textit{“I felt so restless, and really wanted to get rid of it. But there was no way to remove the nasty thing!”} But when verbal cues were disabled, it simply said, \textit{“It could not be removed.”} Importantly, while our animal chatbots’ emotional expressions may resemble those of human-like chatbots, they emphasized its animal identity through contextual framing of these emotions. Notably, most of the emotional expressions occurred during the second segment of the conversation, in which the chatbot narrated experiences that are highly specific to horses.

{\bf Nonverbal Cues.}
As chatbots lack a graphic embodiment, a popular strategy involving nonverbal cues is the use of emojis~\cite{Seeger2018}. We adopted this strategy: i.e., the animal chatbot utilized emojis when nonverbal cues were enabled, but did not use any when such cues were disabled. To distinguish our animal chatbot from those designed to be human-like, its emojis depicted horse faces, such as a smiling horse, a scared horse, and a sad horse (Fig. \ref{fig:emojis}), which we created with the help of Emoji Kitchen~\footnote{https://emojikitchen.dev/}.

{\bf Identity Cues.}
In our animal chatbot, identity cues were implemented through a displayed name and profile picture. In instances where identity cues were enabled, the animal chatbot profile picture was a photograph of a horse (Fig. \ref{fig:chatbot-interface}, left), accompanied by the name {\it "Strider"}. By using a horse profile picture and a name associated with horses, our implementation of identity cues diverged from those typically employed in human-like chatbots. When identity cues were disabled, on the other hand, the chatbot featured a robot as its profile picture (Fig. \ref{fig:chatbot-interface}, right), alongside the name {\it "Horse Bot"}. We chose the name {\it "Strider"} due to its use in the Tolstoy story \textit{Strider: The Story of a Horse} \cite{Tolstoy2015}.

\subsection{Implementation}
\begin{figure}[t]
  \centering
  \includegraphics[width=\textwidth]{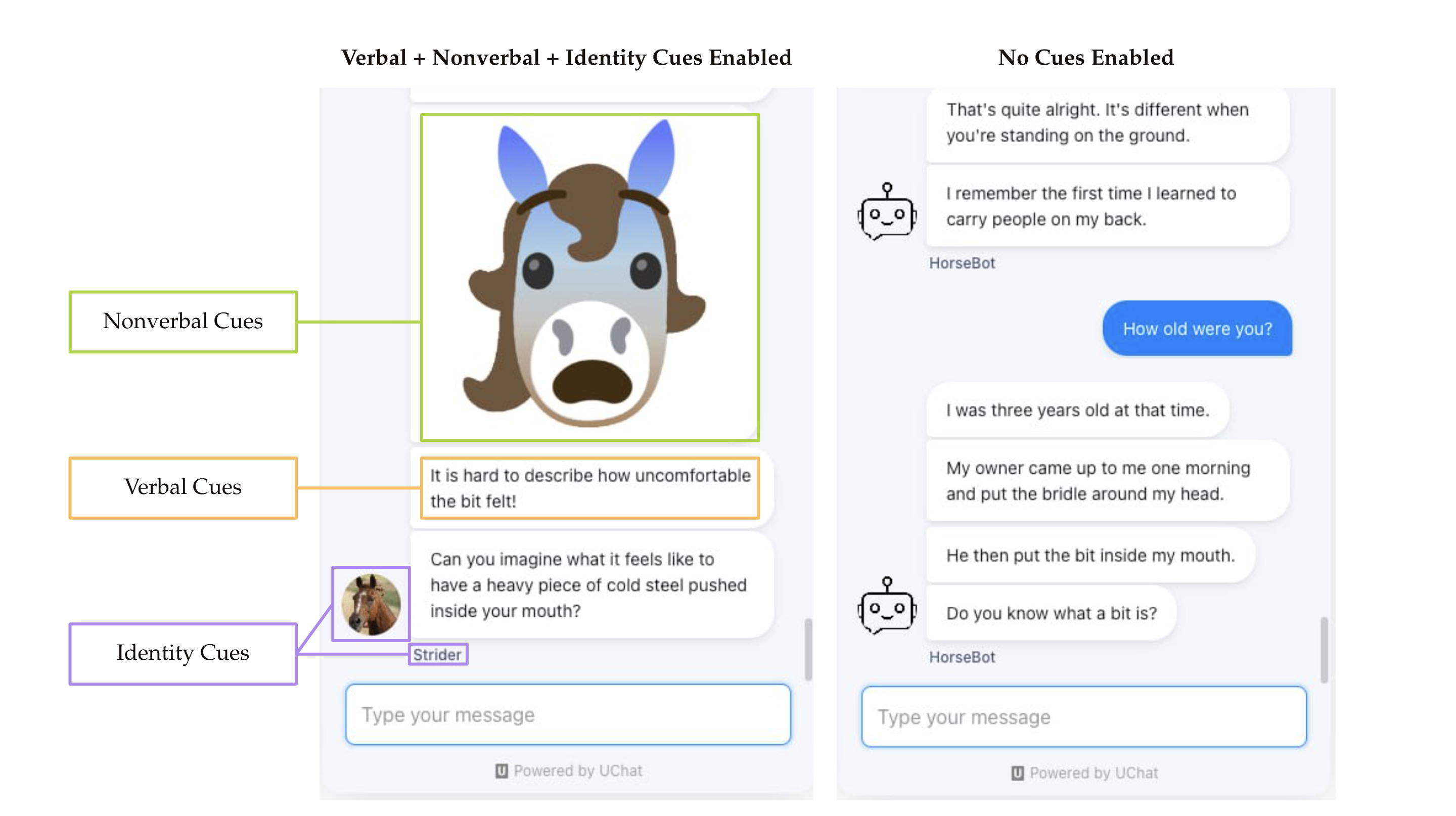}
  \caption{Example screenshots of two versions of the chatbot interface. Left: Verbal, non-verbal, and identity cues enabled. Right: No cues enabled.}
  \label{fig:chatbot-interface}
  \Description{This figure shows the experimental interface in this study. The left side is the conversation interface for the group with verbal, non-verbal and identity cues. The profile photo of the chatbot is a horse. And the chatbot can use emotional verbal expression as well as emoji. The right side is the conversation interface for the group without any cues. The profile photo of the chatbot is a robot. And the chatbot cannot use emoji or emotional verbal expression during conversation.}
\end{figure}

We utilized UChat~\footnote{https://uchat.au/} to develop the animal chatbot. This involved designing a conversational flow capable of dynamically enabling or disabling verbal cues, nonverbal cues, and identity cues. UChat also facilitated the handling of multiple conversations simultaneously; integration with OpenAI, for response generation when required; and real-time conversation tracking, to ensure adherence to the expected flow. Example screenshots of the animal chatbot interface can be seen in Fig. \ref{fig:chatbot-interface}.

To ensure seamless interaction with users while maintaining control over conversation flow, we adopted a hybrid approach that combined rule-based mechanisms and OpenAI~\footnote{https://openai.com/index/openai-api/} GPT-4. Specifically, rule-based mechanisms were used when users responded to the animal chatbot using predefined options, while GPT-4 was employed for generating responses to open-ended user inputs. GPT-4 prompts were tailored based on the enabled cues. 
Below is an example prompt tailored for GPT-4 that guided it to generate contextually appropriate responses consistent with the presence of \textbf{verbal} and \textbf{identity cues}. \textit{"(1) Your name is Strider. (2) You just prompted the user to [imagine the sensation of having a bit shoved into their mouth]. Your task is to craft a suitable response to their reply. (3) You should speak in the perspective of a horse. Be expressive and use emotions wherever appropriate. (4) If the user provides a response that is not relevant to the previous instructions, they should be respectfully declined."}

\begin{figure}[t]
  \centering
  \includegraphics[width=0.8\textwidth]{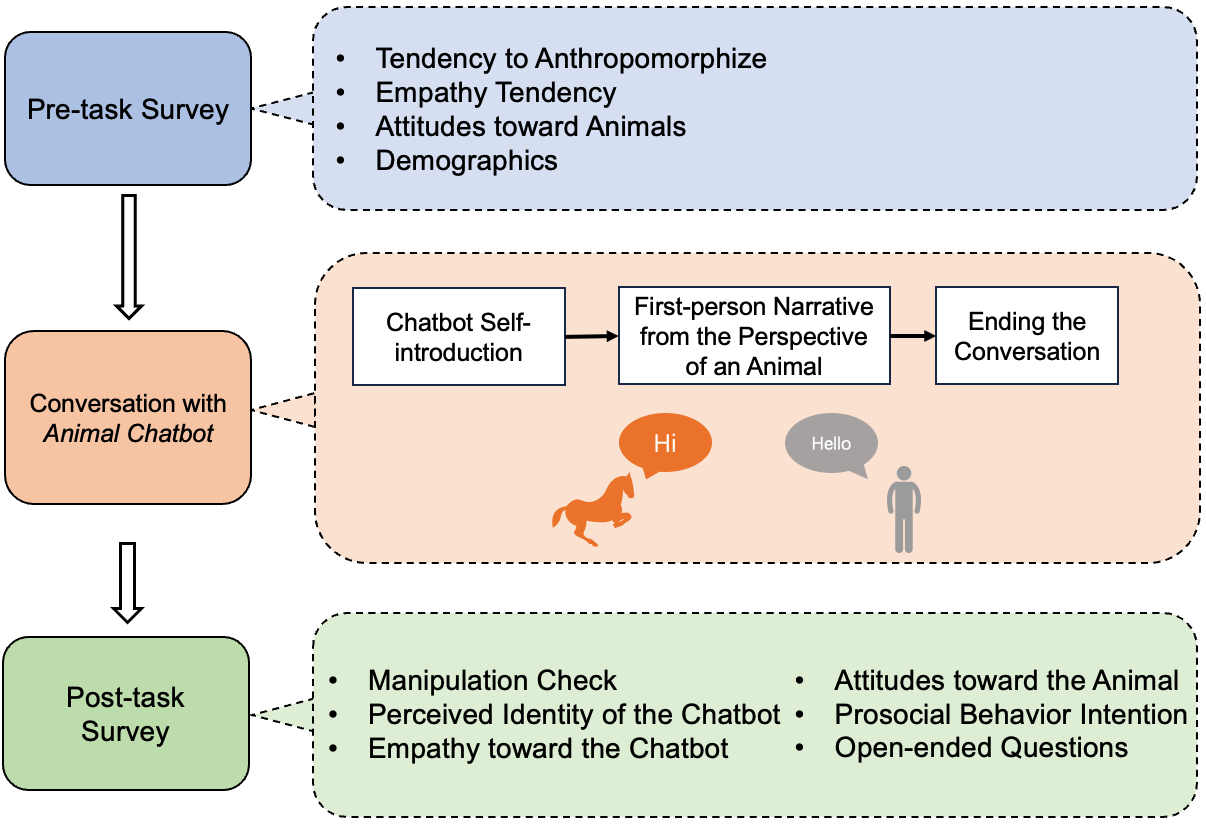}
  \caption{Illustration of the study procedure. Our study includes eight experimental groups according to the combination of cues enabled (verbal cues, nonverbal cues, and identity cues).}
  \label{fig:procedure}
  \Description{This figure shows the experimental procedure of this study. From upper to bottom, there are three phases in order: the pre-task survey, the conversation with animal chatbot, and the post-task survey. In the pre-task survey, controlled variables, including participants' tendency to anthropomorphize, empathy tendency, attitudes towards animals were measured. The demographics of participants were also collected. In the conversation with animal chatbot, the chatbot first introduced itself to participant, then used a first-person narrative to tell its experience about wearing saddles. Finally, the chatbot concluded the conversation and ended it. In the post task survey, manipulation check questions, participants' perceived identity of the chatbot, empathy towards the chatbot, attitudes toward the animal, prosocial behavior intention and open-ended questions were measured.}
\end{figure}

\subsection{Experimental Design}
Given our research focus on the effects of verbal, non-verbal, and identity cues, we employed a between-subjects randomized experiment with a 2 (verbal cues) × 2 (nonverbal cues) × 2 (identity cues) factorial design, in which each cue was treated as an independent binary variable. This resulted in a total of eight experimental groups. The primary dependent variables we assessed were 1) empathy toward the animal chatbot, 2) attitudes toward the animal, and 3) prosocial behavioral intention toward the animal (Fig.~\ref{fig:procedure}).

\subsubsection{Participants}
Our recruitment criteria were that participants must be 1) at least 18 years of age; 2) able to read, write, and speak fluent English; and 3) able to use messaging platforms on their own devices. A total of 240 participants (30 for each condition) were recruited via social-media platforms. The ID of the participant (P) was randomly assigned. In our recruiting poster, we disclosed the duration of the study, along with the participants’ right to drop out at any time. Demographically, 59.2\% identified as female, 39.6\% as male, and 1.2\% as other genders. The majority (94\%) were aged 18-34, and 57\% were educated to bachelor’s-degree level or higher.

\subsubsection{Procedure}
The participants were initially presented with a Participant Information Sheet, which asked them to provide consent before proceeding. Next, they completed a pre-task survey aimed at gathering demographic information and assessing their pre-existing tendencies to empathize and anthropomorphize. They were also notified that if any content or questions made them feel uncomfortable, they could skip them without penalty. Subsequently, they were randomly assigned to one of our eight animal-chatbot versions and instructed to engage with it, in full knowledge that it was not an actual human (i.e., the Wizard of Oz protocol was not being used), and that their responses would not be shared outside the research team prior to anonymization.

Upon completing their respective animal-chatbot conversations, the participants received a password to unlock a post-task survey. This comprised both multiple-choice and open-ended questions designed to gauge their perceptions of and empathy for the chatbot, as well as their attitudes and prosocial behavioral intentions toward the animal it represented. The entire procedure was conducted online for all participants, and we estimated that it would last 15-20 minutes. The participants were compensated US\$4 for their time. Our university’s institutional review board reviewed and approved this research.

\subsection{Instruments}

\subsubsection{Pre-task Survey}

\textbf{Empathy Tendency.} 
Given our interest in the participants’ empathy toward our chatbots, it was crucial to control for their individual empathic tendencies. To achieve this, our pre-task survey incorporated all 28 questions from the Interpersonal Reactivity Index (IRI), which consists of four seven-item subscales: Perspective Taking, Fantasy, Empathic Concern, and Personal Distress \cite{Davis1983}. Each item was rated on a five-point Likert scale ranging from 1=\textit{Strongly disagree} to 5=\textit{Strongly agree}. These questions included: \textit{“Sometimes I don’t feel very sorry for other people when they are having problems”} [reverse-scaled]; \textit{“After seeing a play or movie, I have felt as though I were one of the characters”}; and \textit{“I sometimes feel helpless when I am in the middle of a very emotional situation”}.

\textbf{Tendency to Anthropomorphize.} 
To assess the participants’ individual propensity to anthropomorphize animals and technological devices, we incorporated seven of the 15 questions of the Individual Differences in Anthropomorphism Questionnaire (IDAQ) into the pre-task survey~\cite{Waytz2010}. The eight questions we omitted were deemed irrelevant because they related to the anthropomorphism of natural entities other than animals. Participants rated each IDAQ question on a seven-point Likert scale, ranging from 1=\textit{Not at all} to 7=\textit{Very much}. Three of these questions were: \textit{“To what extent does the average fish have free will?”}, \textit{“To what extent does the average computer have a mind of its own?”}, and \textit{“To what extent do cows have intentions?”}

\textbf{Attitudes toward Animals.} 
As our study aimed to develop a chatbot that can improve attitudes toward animals, it was important to control for participants' pre-existing attitudes toward animals. To accomplish this, we used the {\it Animal Attitude Scale} (AAS-5) during the pre-task survey~\cite{Herzog2015}. Each item on the scale was answered using a five-point Likert scale, ranging from 1 =\textit{Strongly disagree} to 5 = "\textit{Strongly agree}". 
Sample questions from the scale include: "\textit{I sometimes get upset when I see wild animals in cages at zoos}", "\textit{It is morally wrong to hunt wild animals just for sport}", and "\textit{I think it is perfectly acceptable for cattle and hogs to be raised for human consumption}" [reverse-scaled].

\subsubsection{Post-task Survey}
\textbf{Manipulation Check.}
In the post-task survey, we included three manipulation-check questions to ensure that our three independent variables had been successfully manipulated. Regarding the verbal-cues manipulation, participants were asked to rate the emotional richness of the chatbot’s words using a seven-point Likert scale ranging from 1=\textit{No emotions} to 7=\textit{Extremely emotional}. In the case of the nonverbal-cues manipulation, they were prompted to assess the frequency of emoji usage by the agent on a five-point Likert scale ranging from 1=\textit{No emojis} to 5=\textit{Many emojis}. Finally, for the identity-cues manipulation, participants were asked to choose the name of the agent they had interacted with from a list of six options.

\textbf{Perceived Identity of the Chatbot.}
To explore the participants’ perceptions of the chatbot’s identity, our post-task survey posed one question: \textit{“Was the agent you spoke to more like a horse, a human, or a robot?”} Each respondent could select one, two, or all three of the answer options.

\textbf{Empathy toward the Animal Chatbot.}
Empathy can be understood as two distinct processes: \textit{cognitive empathy} and \textit{affective empathy}. Cognitive empathy involves understanding another’s thoughts, feelings, and intentions; i.e., putting oneself in someone else’s shoes and comprehending their perspective, without necessarily feeling what they feel. Affective empathy, on the other hand, involves sharing or mirroring another’s emotional experience; i.e., feeling the emotions they are experiencing, and thereby creating an emotional connection~\cite{Reniers2011}. This bipartite conceptualization of empathy has been applied in prior research on conversational agents~\cite{Liu2018, Park2023}.

Given that cognitive empathy involves recognizing the perspectives of others, our questions about it were similar to those on the Perspective Taking subscale of the IRI~\cite{Davis1983}. Examples included, \textit{“In the conversation, I found it difficult to understand the problems faced by the agent from its point of view”} [reverse-scaled]; \textit{“I believe the agent had experienced pain and distress because of humans”}; and \textit{“In the conversation, I tried to understand the agent’s difficulties by imagining its situation”}. Participants responded to each question using a five-point Likert scale ranging from 1=\textit{Strongly disagree} to 5=\textit{Strongly agree}.

To measure their affective empathy, we asked the participants to rate the extent to which they felt six different emotion-related adjectives applied to themselves after interacting with the animal chatbot, all of which had previously been used in research measuring empathy toward animals~\cite{Batson1987}. They were \textit{sympathetic}, \textit{warm}, \textit{compassionate}, \textit{tender}, \textit{moved}, and \textit{soft-hearted}~\cite{Malecki2019}. Answers were given on a five-point Likert scale ranging from 1=\textit{Not at all} to 5=\textit{Extremely}.

\textbf{Attitudes toward the Animal.}
Given our study’s specific focus on horses, we devised a five-item Horse Attitude Scale. Its questions paralleled those in the AAS-5~\cite{Herzog2015}, which assess people’s tendency to take action to help animals, but were tailored explicitly to horses. Each was rated on a five-point Likert scale, ranging from 1=\textit{Strongly disagree} to 5=\textit{Strongly agree}. The Horse Attitude Scale exhibited good overall reliability (Cronbach’s  $\alpha=0.69$). Its five items were: \textit{“It is perfectly acceptable to hit a horse if they are not following instructions”} [reverse-scaled]; \textit{“It is morally wrong to ride horses just for sport”}; \textit{“It is morally wrong to ride horses as a mode of transportation”}; \textit{“The use of bits to control horses should be stopped if they cause any pain or discomfort”}; and \textit{“It is unethical to consume horse meat”}.

\textbf{Prosocial Behavioral Intention.}
We examined one form of prosocial behavioral intention: willingness to sign a petition~\cite{herrera2018building, taylor2021adolescent}. Specifically, the participants were asked how willing they would be to sign a petition regarding the use of horse tack, which they answered on a five-point Likert scale ranging from 1=\textit{Strongly disagree} to 5=\textit{Strongly agree}.

\subsubsection{Qualitative Measurements}
To gain a deeper understanding of how participants perceived our animal chatbots, we included open-ended questions in the post-task survey. These mainly focused on 1) how the participants perceived the identity of their animal chatbot, and why; 2) whether/how the animal chatbot affected their emotions and attitudes, and why; and 3) their impressions of and attitudes toward horseback riding, and why.

To analyze the responses to these questions, we used thematic analysis. Initially, two researchers independently familiarized themselves with and assessed the responses of the same subsample of 30 participants, chosen at random from all eight chatbot conditions. Those two researchers achieved a strong Cohen’s kappa score for inter-rater reliability, i.e., > 0.8. Subsequently, divergent ratings for the same 30 participants were discussed until a consensus was reached. Following this calibration phase, the remaining responses were evenly distributed between the two raters, with each evaluating the responses of an additional 105 participants.

\section{Results}
An analysis of variance (ANOVA) test of the pre-task survey data indicated that there were no significant differences among the eight experimental groups in terms of their members’ individual tendencies to anthropomorphize animals ($F(1, 232) = 0.559$, $p = 0.789$) or machines ($F(1, 232) = 0.724$, $p = 0.652$). 
Similarly, there were no significant inter-group variations in attitudes toward animals ($F(1, 232) = 0.395$, $p = 0.905$). 
No significant differences were observed for three of the four IRI subscales, which were used to measure tendencies to empathize. These three were Empathic Concern ($F = 0.139$, $p = 0.995$), Perspective Taking ($F(1, 232) = 1.038$, $p = 0.405$), and Personal Distress ($F = 0.580$, $p = 0.771$). However, significant inter-group differences were found in Fantasy Scale scores ($F(1, 232) = 2.168$, $p < 0.05$), which were therefore used as a control variable in all subsequent quantitative analyses.

\subsection{Perceptions of the Chatbot's Identity (RQ1)}
\subsubsection{Manipulation Check}
A one-way analysis of covariance (ANCOVA) test revealed that participants from the four groups whose chatbots used verbal cues ($M = 5.433$, $SD = 0.976$) perceived their chatbots’ language to be significantly more emotionally rich ($F(1, 231) = 66.806$, $p < 0.001$) than participants from the other four groups did ($M = 4.075$, $SD = 1.518$). Participants from the four groups that experienced nonverbal cues ($M = 4.192$, $SD = 0.781$) perceived their chatbots’ use of emojis to be significantly more frequent ($F(1, 231) = 1498.484$, $p < 0.001$) than those from groups that did not experience such cues ($M = 1.058$, $SD = 0.416$). Lastly, almost all (98.8\%) of the participants correctly identified the name of the chatbot with which they had interacted. Hence, we concluded that the manipulations of all three independent variables were successful.

\subsubsection{Perceived Chatbot Identity}
To answer RQ1, we analyzed the participants’ responses for evidence of whether they perceived the identity of their chatbot to be a bot, a human, or an animal, as this information would help us infer the influence of verbal, nonverbal, and identity cues. The results were fairly evenly split, with 86 participants perceiving their chatbots as robot-like, 54 as horse-like, and 82 as human-like. Among the remaining 18 participants, four perceived their chatbots as robot-like + horse-like; six as robot-like + human-like; four as horse-like + human-like; and four as robot-like + horse-like + human-like.

We employed multiple chi-square tests to evaluate how these perceptions of chatbot identity related to our three different designs of cues. We found a significant relation between verbal cues and participants’ perceptions of chatbot identity ($\chi^2(2, N=222)=11.186$, $p<0.01$). That is, when verbal cues were present, users were more likely to perceive their chatbot as human-like than when verbal cues were absent. The effects of non-verbal cues or identity cues were not significant.

\subsubsection{Qualitative Findings}
While every participant was aware that they were communicating with a chatbot, their answers to open-ended questions suggest varied perceptions of the chatbot’s identity. When asked how they perceived the animal chatbot’s identity, around one-third of the 120 participants that received verbal cues (n = 45) noted that there was no discernible difference between the chatbot’s speech and that of a human, because emotive words like \textit{“uncomfortable”} and \textit{“pain”} were used. As P232 stated, \textit{“The chatbot was conveying emotions and words that we would expect humans to say. The only difference is in the context of being a horse.”}

However, around one-fifth of the participants in the verbal-cues groups (n = 21) felt that the animal chatbot spoke authentically about the experiences, emotions, and scenarios typical of a horse’s life. This perspective made the interaction feel more like talking to a horse than to a human or robot. As P48 noted, \textit{“The agent I spoke to felt more like a horse, as it was able to describe the story from a horse’s point of view. It also managed to explain the feelings of the horse in the story accurately, or as I would think a horse feels in such a situation.”} In addition, the chatbot’s use of horse-related terms and contexts (e.g., being ridden, wearing a saddle) reinforced its horse-like identity. P139 told us: \textit{“It was more of a horse. He started talking about how he was raised in a farm, setting up the context of a farm animal \& how he was first being ridden by a human (putting on the bridle, bit \& saddle) that he felt uncomfortable \& nervous.”}

Several other participants from the verbal-cues groups (n = 6) expressed mixed feelings about the chatbot’s identity, perceiving it to be somewhere between animal and human. While horses do not speak, the perspective and life experiences conveyed by the horse-chatbot made it feel authentic to these users; and they also felt that the chatbot’s communication style was emotionally expressive and relatable in a way that resembled human interaction. For example, P51 mentioned that \textit{“[t]he agent I talked to seemed more like a mix of a human and a horse in expressing emotions. It displayed feelings similar to a horse but in a way that humans can understand, creating a blend of both characteristics.”}

Conversely, in the four groups whose chatbots lacked verbal cues, nearly half the participants (n = 51) stated that their chatbots seemed robotic. They attributed this perception to their bots’ strictly factual retelling of events and their lack of expression of emotion or personal thoughts. Although the conversation had a logical progression, it lacked the genuineness of human interactions. For example, P169 remarked: \textit{“[H]orses can’t talk but if they could, I’m quite sure they would have feelings and preferences that they would express to me, unlike the agent who used plain words that didn’t contain any emotion. Definitely didn’t sound like a human as humans would have more emotions in their words especially if they were describing a personal experience.”}

Within the four groups where nonverbal cues appeared in the conversation, some participants (n = 17) felt that the chatbot resembled a human because the emotions conveyed through the emojis were quite similar to those of humans. As P28 mentioned, \textit{“[t]he frequent use of emojis to express its emotions was a bit goofy but endearing. Being able to express conscious thoughts coherently just made me think it was more like a human.”}

However, a few participants in the nonverbal-cues groups (n = 13) reported that the use of emojis made the chatbot feel artificial, or \textit{“very synthetic”}, as P230 put it. The same participant elaborated that \textit{“I felt like the agent spoke very matter-of-factly [...] and it was weirdly interspersed with emojis that felt like someone else put them there.”}

Interestingly, most of the participants who encountered identity cues during their animal-chatbot interactions did not consider them to be particularly important. However, three participants noted that such cues made the chatbot seem horse-like to them. As P223 explained, \textit{“The agent felt more like a horse to me as [… it] was using first-person language all the time and also of the display picture of a horse. Also, the name of the agent is Strider which sounds more like a horse name rather than a robot name and it does not sound like a human name at all.”}

\subsection{Effect of Chatbot Design Cues (RQ2)}

\begin{figure}[t]
  \centering
  \includegraphics[width=0.7\textwidth]{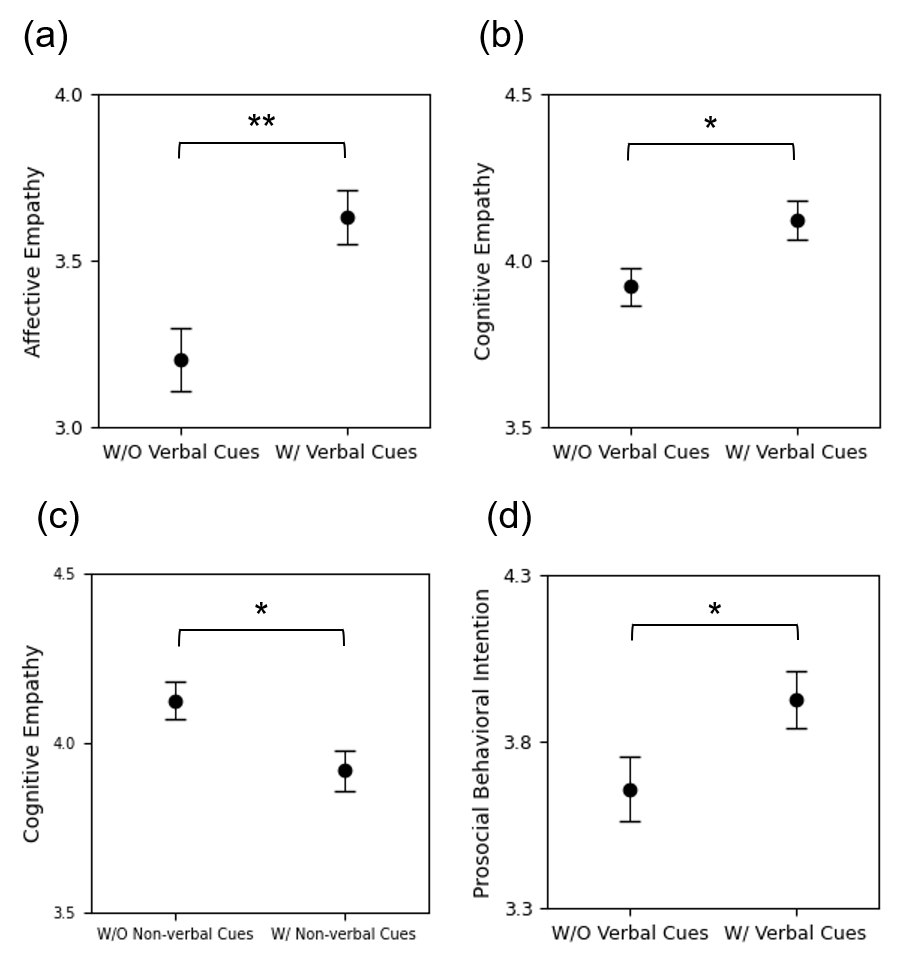}
  \caption{Results about the effects of chatbot design cues. The dots represent mean values, and the error bars show one standard error. The significance levels are labeled ($p<0.05$: *, $p<0.01$: **, $p<0.001$: ***).
  {\bf (a)}: Effect of verbal cues on affective empathy.
  {\bf (b)}: Effect of verbal cues on cognitive empathy.
  {\bf (c)}: Effect of {\bf non-verbal} cues on cognitive empathy.
  {\bf (d)}: Effect of verbal cues on prosocial behavior intention.
  }
  \label{fig:anthro factors effects}
  \Description{This figure shows the results about the effects of chatbot design cues. There are four subfigures, representing effect of verbal cues on affective empathy, effect of verbal cues on cognitive empathy, effect of non-verbal cues on cognitive empathy, and effect of verbal cues on prosocial behavior intention. The dots represent mean values, and the error bars show one standard error. The significance levels are labeled ($p<0.05$: *, $p<0.01$: **, $p<0.001$: ***).}
\end{figure}

\subsubsection{Effect on Empathy}
We conducted a three-way ANCOVA to assess the impacts of verbal, nonverbal, and identity cues on cognitive and affective empathy. We observed a significant main effect of verbal cues on affective empathy ($F(1, 231) = 9.766$, $p < 0.01$, $\eta^2=0.038$), with participants from groups with verbal cues ($M = 3.629$, $SD = 0.885$) demonstrating significantly higher levels of affective empathy than others ($M = 3.204$, $SD = 1.031$), as shown in Fig.~\ref{fig:anthro factors effects}(a). This result indicates that verbal cues played a significant role in enhancing affective empathy.

Significant main effects of both verbal cues ($F(1, 231) = 4.741$, $p < 0.05$, $\eta^2=0.019$) and nonverbal cues ($F(1, 231) = 6.482$, $p < 0.05$, $\eta^2=0.026$) on cognitive empathy were also observed. Further analysis indicated that while participants from groups with verbal cues ($M = 4.122$, $SD = 0.658$) exhibited higher levels of cognitive empathy than others ($M = 3.922$, $SD = 0.605$), as shown in Fig.~\ref{fig:anthro factors effects}(b). Those from groups with nonverbal cues ($M = 3.919$, $SD = 0.660$) displayed lower levels of cognitive empathy than others ($M = 4.125$, $SD = 0.602$), as shown in Fig.~\ref{fig:anthro factors effects}(c). These results suggest that verbal and nonverbal signals both influenced cognitive empathy, but in opposite directions.

\subsubsection{Effect on Attitude} 

We conducted a three-way ANCOVA to investigate the impact of verbal, nonverbal, and identity cues on the participants’ attitudes toward horses. The results indicated that none of these chatbot-design cues had a significant effect on such attitudes.

\subsubsection{Effect on Prosocial Behavioral Intention}
Another three-way ANCOVA unveiled a significant main effect of verbal cues on prosocial behavioral intention, i.e., willingness to sign a petition regarding the use of horse tack ($F(1,231) = 4.034$, $p <0.05$, $\eta^2=0.017$). Specifically, as shown in Fig.~\ref{fig:anthro factors effects}(d), participants from the groups with verbal cues ($M = 3.925$, $SD = 0.936$) expressed a significantly higher willingness to sign than other participants did ($M = 3.658$, $SD = 1.049$).

\subsubsection{Qualitative Findings}
Our analysis of the open-ended question also suggested that verbal cues played an important role in eliciting empathy. Most of the participants (n = 96) indicated that they were emotionally touched as the chatbot vividly recounted its painful experiences and emotions, and that this allowed them to envision riding a horse from the horse’s point of view. As P57 noted, \textit{“The words and phrases they used were very human-like, [and] evoked feelings of compassion and empathy in me. It gives the perspective of a horse that humans are unable to understand.”} P232 further emphasized the impact of the painful experiences that the chatbot shared: \textit{“I feel empathetic about the horse’s experience and it felt […] painful and unpleasant”}.

In contrast, many participants (n = 47) in the four conditions without verbal cues told us that they were not emotionally affected by the chatbot and highlighted their chatbots’ limited ability to convey emotions during interactions, which negatively affected their ability to empathize with them. As P116 mentioned, \textit{“I did not feel moved. He talked about his experience but he didn’t describe his feelings or how hurt he was. He just stated he was scratched, no description or emotions were talked about.”}

Interestingly, we found that the use of emojis as nonverbal cues was a double-edged sword when it came to evoking participants’ emotions. Some participants (n = 26) mentioned that emojis improved their sense of emotional connection to the animal chatbot. P132, for instance, said: \textit{“I feel moved by Horsebot during the conversation, as it sounds like it spoke from real experience. It was able to describe its painful experience in great detail, and [this] made it very relatable. The use of emojis was very appropriate too and enhanced the moving encounter.”} 
However, others (n = 7) mentioned that emojis detracted from the seriousness or genuineness of the conversation, and thus negatively impacted their sense of empathy toward the animal chatbot. For instance, P152 noted, \textit{“There was this touch of sympathy when the convo got going. But the emoticons made it less serious.”}

\subsection{Effect of Perceived Chatbot Identity (RQ3)}

We excluded the responses from the above-mentioned 18 participants who perceived their chatbots as having multiple identities, on the grounds that we were unable to categorize them into single types. Therefore, the data on which we based our answer to RQ3 was drawn from the 222 participants who perceived a single type of identity, i.e., only robot-like (n = 86), only horse-like (n = 54), and only human-like (n = 82).

\begin{figure}[t]
  \centering
  \includegraphics[width=0.7\textwidth]{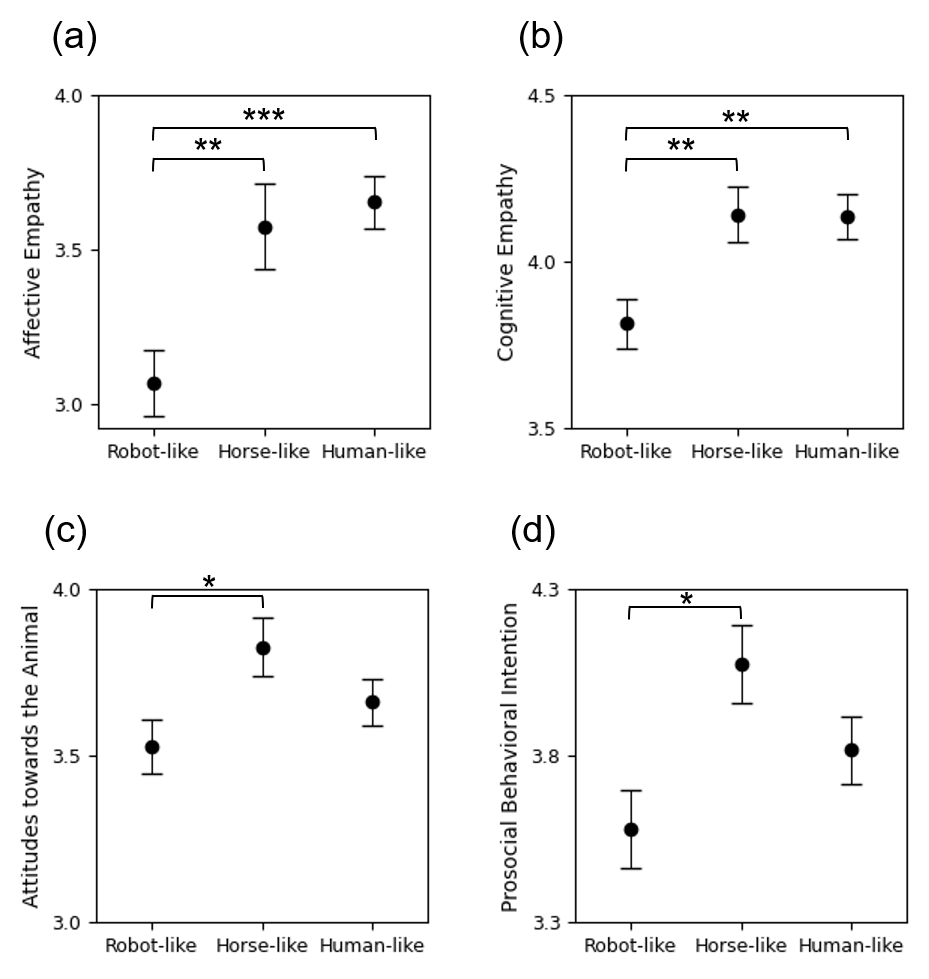}
  \caption{Results about the effects of perceived identity the chatbot. The dots represent mean values, and the error bars show one standard error. The significance levels are labeled ($p<0.05$: *, $p<0.01$: **, $p<0.001$: ***).
  {\bf (a)}: Effect on affective empathy.
  {\bf (b)}: Effect on cognitive empathy.
  {\bf (c)}: Effect on attitudes toward the animal.
  {\bf (d)}: Effect on prosocial behavior intention.
  }
  \label{fig:perceived identity}
  \Description{This figure shows the results about the effects of perceived identity the chatbot. There are four subfigures, representing effect on affective empathy, effect on cognitive empathy, effect on attitudes towards the animal, and effect on prosocial behavior intention accordingly. The dots represent mean values, and the error bars show one standard error. The significance levels are labeled ($p<0.05$: *, $p<0.01$: **, $p<0.001$: ***).}
\end{figure}

\subsubsection{Effect on Empathy}

We performed a one-way ANCOVA to explore the impact of the three categories of perceived chatbot identity on participants’ cognitive and affective empathy. This was followed by a post-hoc analysis using Tukey’s honestly significant difference (HSD) test, which compares pairwise means with corrections for multiple comparisons. There was a significant main effect of perceived chatbot identity ($F(2, 218) = 10.244$, $p < 0.001$, $\eta^2=0.083$) on affective empathy, as shown in Fig.~\ref{fig:perceived identity}(a). Post-hoc analysis ($t=3.460$, $p<0.01$, $d=0.604$) indicated that those participants who perceived their bots as horse-like ($M=3.574$, $SD=1.021$) exhibited a significantly higher level of affective empathy than those who perceived theirs as robot-like ($M=3.068$, $SD=1.003$). The group that perceived their bots as human-like ($M=3.654$, $SD=0.770$) also had a significantly ($t=4.123$, $p<0.001$, $d=0.636$) higher level of affective empathy for them than those who perceived their bots as robot-like. However, no significant difference in affective empathy was found between the participants who perceived their bots as horse-like and those who found them human-like ($t=-0.187$, $p=0.981$, $d=-0.033$).

In addition, there was a main effect of perceived chatbot identity ($F(2, 218) = 7.555$, $p < 0.001$, $\eta^2=0.062$) on cognitive empathy, as shown in Fig.~\ref{fig:perceived identity}(b). Post-hoc analysis showed that the participants who perceived their chatbots as horse-like ($M=4.142$, $SD=0.606$) exhibited a significantly higher level of cognitive empathy ($t=3.295$, $p<0.01$, $d=0.575$) than those who perceived them as robot-like ($M=3.814$, $SD=0.670$). Those who perceived their bots as human-like ($M=4.134$, $SD=0.617$) also were found to have a significantly higher level of cognitive empathy ($t=3.280$, $p<0.01$, $d=0.506$) than those who perceived their bots as robot-like. But again, no significant difference ($t=-0.388$, $p=0.921$, $d=0.068$) in cognitive empathy was observed between the groups that perceived their bots as horse-like vs. human-like. These results indicate that perceived chatbot identity significantly impacted both affective and cognitive empathy toward the horse chatbot, with a robot-like identity eliciting the lowest levels of both types of empathy.

\subsubsection{Effect on Attitude}
The one-way ANCOVA we conducted to investigate the impact of perceived chatbot identity on attitudes toward horses found a significant main effect ($F(2, 218) = 3.460$, $p < 0.05$, $\eta^2=0.030$), as shown in Fig.~\ref{fig:perceived identity}(c). Specifically, those participants who perceived their chatbots as horse-like ($M = 3.826$, $SD = 0.647$) expressed significantly more positive attitudes toward horses than those who perceived them as robot-like ($t=2.628$,$p<0.05$, $d=0.458$; $M = 3.528$, $SD = 0.759$). However, no significant differences were observed between the “horse-like” and “human-like” groups ($t=1.519$, $p=0.284$, $d=0.268$) or between the “human-like” and “robot-like” groups ($t=1.235$, $p=0.434$, $d=0.191$). The results indicate that participants who perceived their chatbots as horse-like had significantly more positive attitudes toward horses compared to those who perceived their chatbots as robot-like.

\subsubsection{Effect on Prosocial Behavioral Intention}
A one-way ANCOVA unveiled a significant main effect of perceived chatbot identity on prosocial behavioral intention ($F(2, 218) = 4.352$, $p < 0.05$, $\eta^2=0.038$), as shown in Fig.~\ref{fig:perceived identity}(d). Specifically, those participants who perceived their chatbots as horse-like ($M = 4.074$, $SD = 0.866$) expressed significantly higher willingness to sign a petition than those who perceived theirs as robot-like ($t=2.931$, $p < 0.05$, $d=0.511$; $M = 3.581$, $SD = 1.079$)). However, no significant differences were observed between the “horse-like” and “human-like” groups ($t=1.532$, $p=0.278$, $d=0.270$) or between the “human-like” and “robot-like” groups ($t=1.5632$, $p=0.264$, $d=0.241$). These results suggest that participants who saw their chatbots as horse-like showed significantly higher intention to sign a petition than those who saw theirs as robot-like.

\subsubsection{Qualitative Findings}
Several key themes emerged from our analysis of the participants’ responses about how chatbot identity affected their perceptions. Many participants (n = 48) who perceived their chatbot as human-like expressed a decreased willingness to ride horses. As P95 put it, \textit{“Not willing, because I will recall this conversation with the agent. That the horse has to endure my body weight and pain of having to carry me through a long distance.”}

Several participants (n = 15) who perceived the animal chatbot as human-like found its painful experiences and emotional expressions somewhat moving, but their awareness of its true nature as a chatbot brought them back to reality, hindering deeper emotional engagement. As P208 noted, \textit{“I needed to imagine the experiences of the horse and listen to its experience of being abused as though it was a human. But it was a bot and created by humans, so I was limited [… in] sympathy to the horse.”}

Most (n = 31) of the participants who perceived the animal chatbot as horse-like indicated that the authentic portrayal of the life of a horse resonated with them. They said that they felt sympathetic toward its plight, and that they believed the emotions and scenarios were true to a horse’s experience. This, in turn, made them uncomfortable with the idea of horseback riding, by increasing their awareness of its ethical and physical implications. P72 said, \textit{“I feel sorry for the agent experiencing those circumstances and, knowing its story as a horse, I felt uneasy about the fact that there are horses out there experiencing similar situations. The agent made me realize that horses do not enjoy the process of being trained for human riding, as they have a conscious mind and are aware of their feelings.”}

Many participants who perceived the horse chatbot as robot-like (n=50) had the lowest levels of emotional engagement with it, as they found it hard to connect with a being they knew was artificial. P187 explained: \textit{“I did not feel emotionally affected during or after the conversation with the agent because I was aware that I was interacting with a bot that was programmed to respond to me in that manner.”} 
These participants tended to regard horseback riding in a detached manner – e.g., focused on the practical uses of horses throughout history – and maintained that as long as horses are not mistreated, riding them is acceptable. Their lack of emotional engagement arising from the chatbot’s robotic nature made it easier for them to justify their continued willingness to ride. P218 told us that, although horses should not be hit, they \textit{“are useful and are essential in our daily lives. Before cars were invented, we had to rely on horses to get around quickly, [and] had we not used them, it would slow our progress overall for quite a bit. I am alright with using horses because it doesn’t seem to be as harmful to them.”}

\section{Discussion} 
\subsection{Impact of Verbal Cues on the Perceived Identity of Chatbots (RQ1)}
Our research builds on previous studies that highlighted the role of verbal cues in shaping users’ perceptions of chatbot identity. Our findings echo those of prior work that found specific verbal cues were able to increase chatbots’ human-like qualities~\cite{Seeger2018, Liu2018}. We have extended this understanding by illustrating how these cues can also encourage perceptions of a chatbot as having an animal identity. For example, verbal cues that authentically conveyed a story and emotions resulted in some users perceiving a chatbot as horse-like. This is intriguing, insofar as animals do not speak; and people’s notion of \textit{“authenticity"} in connection with animal-like entities warrants further exploration.

Our first-person narrative from a horse’s perspective, inspired by Anna Sewell’s Black Beauty \cite{Sewell1877}, underscores the potential of leveraging classic literature to explore the theoretical underpinnings of chatbot design. This approach could pave the way for more theory-driven investigations into how chatbots might authentically represent various non-human identities. Such explorations could consider multiple dimensions of authenticity, including authenticity of source, of message, and of interaction~\cite{lee2020authenticity, huang2022perceived}.

Specifically, chatbot designers could potentially enhance authenticity of interaction – i.e., people’s feeling that they are part of a genuine interaction – by engaging users in reciprocal interaction during storytelling~\cite{liang2024dialoging, lee2019caring} to build a sense of mutual awareness, attention, and adjustment \cite{hartmann2011horton}. However, because humans lack a common language with animals, it is still challenging to develop message authenticity, i.e., narratives that are true representations of their subject matter~\cite{lee2020authenticity, huang2022perceived}. Describing animals based on their bodily movements might be a means of enhancing message authenticity, in light of Aspling et al.’s ~\cite{aspling2018understanding} findings regarding \textit{kinesthetic empathy}, i.e., empathizing with an animal by understanding their bodily experiences over time and their individual histories. Further research should delve deeper into these aspects and refine our understanding of the complex interplay between verbal cues and perceived authenticity in both human-like and non-human-like chatbot identities.

\subsection{Impact of Verbal and Nonverbal Cues on Empathy in Animal Chatbots (RQ2)} 
Our quantitative findings demonstrate that verbal cues significantly enhanced both cognitive and affective empathy with our animal chatbots. The qualitative data support this, and further reveal that exposure to verbal cues helped participants build a deeper emotional connection with the chatbot’s narratives of pain and various emotions. This increase in empathy echoes prior literature on animal narratives, which stresses the importance of human-like emotions to fostering empathetic responses~\cite{Poirier2019, Young2018}. Our results are also consistent with previous research that highlighted how chatbots expressing human-like emotions can elicit user compassion~\cite{lee2019caring}.

Conversely, our investigation into nonverbal cues revealed a surprising decrease in the participants’ cognitive empathy when emojis were used – in contrast to prior studies that found emojis enhanced the credibility and warmth of human-like chatbots, thereby boosting cognitive empathy~\cite{Yu2024, Beattie2020}. There are two likely explanations for this discrepancy. First, the emojis used in our study depicted horse faces, which might have been more challenging to interpret than the human-face emojis used in earlier research. Qualitative feedback indicated that these emojis were perceived as artificial and relatively unexpressive, complicating users’ ability to see things from the chatbot’s perspective. Second, participants’ reactions to the emojis were mixed. Some found them appropriate and evocative, while others – perceiving them as humorous rather than serious – saw them as trivializing the discussion. This mixed reception likely reduced the overall perceived gravity of the chatbot narrative, thus lowering empathy levels. Prior research has shown that while emojis can sometimes enhance relational communication in negative contexts, they are often deemed inappropriate for such contexts~\cite{cavalheiro2022evaluating, rodrigues2017frown}. Future research could therefore usefully compare the effects of human vs. non-human face emojis in serious/negative contexts.

\subsection{Enhancing Empathy and Prosocial Attitudes through Perceptions of Animal-like Chatbots (RQ3)}
Previous research has highlighted how emotionally engaging experiences can effectively foster empathy toward animals and the natural world, and suggested that interactions involving humans, animals, and computers can catalyze this empathy by enhancing animals’ similarities to humans~\cite{webber2017kinecting, webber2023engaging}. In line with those studies’ findings, our participants who perceived our animal-chatbot as human-like demonstrated greater empathy toward it than those who saw it as robot-like. This also echoes prior findings that attributing human-like qualities to digital agents can increase their users’ empathetic responses~\cite{lee2019caring, araujo2018living}.

Our results also provide new insights into the role of animal-likeness perceptions in human-chatbot interactions. Participants who perceived the animal chatbot as horse-like not only expressed higher empathy toward it but also displayed more positive attitudes and higher prosocial behavioral intention toward actual horses than those who viewed it as robot-like. This may mean that chatbots widely and strongly perceived as animal-like could boost people’s aversion to animal cruelty. Such an effect might be driven by two factors. The first was the perceived authenticity \cite{curtin2005nature} of the chatbot’s portrayal of a horse’s life, which strongly influenced participants’ attitudes toward horseback riding. Second, while direct studies of animal-like chatbots are scarce, the literature on animal-like robots suggests that they can enhance engagement and heighten perceptions of animals’ social intelligence and warmth \cite{miklosi2012utilization}, and thus increase pro-conservation attitudes and behaviors. Our study extends that literature by demonstrating that not only robots, but also chatbots designed to represent non-human identities can effectively influence human attitudes and behavior intentions toward those entities.

Nevertheless, further research is needed to explore how chatbots can represent non-human entities authentically \cite{huang2022perceived}, as a means of engaging people with nature and promoting conservationist attitudes. Such investigations could pave the way for theory-driven approaches to chatbot design aimed at altering behaviors and attitudes, and potentially have an important impact on real-world conservation efforts.

\subsection{Design Implications}
Chatbots have demonstrated their potential to support self-improvement on the part of their users, e.g., by mitigating their public-speaking anxiety; and a key design strategy for achieving such goals is to equip chatbots with human-like communicative behaviors~\cite{van2020human}. The present study’s findings have explored the similar, but distinct approach of designing chatbots to represent non-human identities. We found that this could foster empathy and positive attitudes toward the non-human entities they represented. By employing first-person narratives from a non-human perspective~\cite{yoneyama2021miyazaki, Malecki2019}, chatbots can effectively convey environmental concerns and animal stories, encouraging individuals to reconsider their relationship with nature and other beings. While non-human first-person narratives are not novel, their application in chatbots offers a unique opportunity to engage audiences with animal-welfare and other environmental/conservation issues.

Because the design of our animal chatbot was inspired by prior techniques aimed at making chatbots more human-like, 42\% of our participants (n = 100) – perhaps inevitably – either deemed it to be human-like or expressed uncertainty about its identity. Future researchers could therefore usefully revisit the effectiveness of verbal cues, using a variety of techniques to effectively convey the emotions and perspectives of non-human entities in their narratives. For example, in addition to expression of emotions, Bernaerts et al.~\cite{bernaerts2014storied} recommended multiple techniques for non-human storytelling, such as highlighting the differences between human and non-human experiential states, to deepen the impact of non-human perspectives. By contrasting these experiences against typical human ones, designers should be able to defamiliarize readers with their normal ways of seeing the world, and thus enhance their narratives’ empathy-building and thought-provoking qualities.

To our surprise, nonverbal cues consisting of animal-face emojis had a negative impact on empathy. This highlights the necessity of being cautious when integrating such elements into animal chatbots intended to elicit empathy. Further research should seek ways to use emojis without undermining the seriousness of the discourse. Specifically, while animal-face emojis might reduce empathy by making messages seem less serious, Riordan et al.~\cite{riordan2017communicative} have reported that non-face emojis can minimize ambiguity in messages and convey emotional content. This could be especially valuable in computer-mediated communication that lacks the usual nonverbal cues of face-to-face communication.

Interestingly, identity cues did not influence overall user perceptions of our chatbots, regardless of whether they had bot-like or horse-like traits. This contrasts with prior findings that presenting a human identity usually impacts how users perceive chatbots \cite{Go2019}. This discrepancy could have been due to the innate implausibility of chatting with a horse. Future studies should therefore examine other forms of identity cues, like using embodied conversational agents with animal identities~\cite{Byrne2012, Chen2011}. Additionally, animal-like companion robots (e.g., Paro - seal-like companion robot~\cite{miklosi2012utilization}) could be assessed for their ability to boost their users’ views of conversational agents as animal-like.

\subsection{Limitations and Future Research}
This paper has several limitations that could be addressed in future research. First, its exploration was limited to specific forms of verbal, nonverbal, and identity cues. In particular, it focused on the expression of emotions as verbal cues, the use of horse emojis as nonverbal cues, and the use of a horse’s name and profile picture as identity cues. Future studies could therefore usefully explore the effects of alternative forms of verbal, nonverbal, or identity cues on animal chatbots’ effectiveness.

Second, our chatbots were developed to discuss the specific topic of horseback riding. Future research could enrich this paper’s findings by exploring chatbots that discuss other topics relating to horses, such as the consumption of horse meat, and/or by developing chatbots that represent other animals, ranging from pets like cats to ones typically viewed as pests.

Third, our study only required its participants to specify if they perceived the chatbot’s identity as human-like, horse-like, and/or robot-like, instead of rating these identities on a numerical scale. This was due to the exploratory nature of our research and the lack of any established definition or measurement scale for animal-like characteristics. Therefore, our goal was to investigate how people described their assessment of a bot as “horse-like” in qualitative terms. Based on those results, future researchers could delve deeper into this topic.

Lastly, our research design did not include any experimental groups that received the same horse narrative from non-chatbot media such as books or documentary films. In future, such comparisons could provide valuable information on the unique advantages and limitations of chatbots as a medium to convey information and influence perceptions.

\section{Conclusion} 
This study developed a chatbot that speaks from the perspective of a horse and examined its impact on users’ perceptions of horses. Its findings demonstrate that emotional expressions (verbal cues) and the perceived authenticity of the animal’s identity both evoked empathy for the chatbots and perceptions of them as horse-like. Authenticity emerged as a key factor, with narratives that convincingly conveyed a horse’s perspective significantly increasing user empathy toward both the chatbot and real horses. Conversely, the use of nonverbal cues, specifically animal-face emojis, reduced cognitive empathy due to their perceived artificiality; while identity cues had hardly any impact. These insights underscore the potential of animal-like chatbots to foster empathy and positive attitudes toward animals and the environment, while also highlighting the need for careful selection of nonverbal elements to maintain the authenticity and seriousness of the discourse. Future research should delve deeper into the complexities of animal-chatbot design, with a focus on developing authentic non-human identities that can engage users in meaningful, empathy-evoking interactions.


\bibliographystyle{ACM-Reference-Format}
\bibliography{sample-base}


\appendix
\clearpage
\section{Chatbot Scripts, Figures, and Prompts}
Here are the chatbot scripts of two conditions: (verbal cues, non-verbal cues, identity cues), (non-verbal cues). For the condition without non-verbal cues, simply remove all emojis. All responses generated by GPT are marked as GPT-generated, while other responses are scripted. All emojis used in the scripts and the photo of Strider are provided in section \ref{sec:emoji}.

\subsection{Script with Verbal Cues, Non-verbal Cues and Identity Cues}
\begin{itemize}
    \item {\bf Strider}: Hello! My name is Strider. <Emoji 1> 
    \item {\bf Strider}: How may I address you? You do not have to give me your real name.
    \item {\bf Participant}: <Participant Name>
    \item {\bf Strider (GPT-generated)}:  <Participant Name>, nice to see you! 
    \item {\bf Strider}: Let me share a bit more about myself!
    \item {\bf Strider}: I am a horse, and I live with my owner and many other animals on a farm. 
    \item {\bf Strider}: <Strider's Photo>. Here is a picture of me!
    \item {\bf Strider}: What about you, do you also live and work on a farm?
    \item {\bf Participant}: No.
    \item {\bf Strider}: That is alright! Working on a farm can be really tiring and painful sometimes.
    \item {\bf Strider}: Every day, I am supposed to carry my owner around the farm and help him with his work.
    \item {\bf Strider}: While doing that, I also have to wear a lot of things, such as a bridle, bit, and saddle. <Emoji 2> 
    \item {\bf Strider}: I remember the first time I learned how to do all of this. It was such a painful experience…
    \item {\bf Participant}: How old were you?
    \item {\bf Strider}: I think I was only three years old at that time.
    \item {\bf Strider}: I remember my owner coming up to me one morning and putting the bridle around my head.
    \item {\bf Strider}: Before I could react, he pushed the bit inside my mouth.  <Emoji 3> 
    \item {\bf Strider}: It is hard to describe how uncomfortable the bit felt!
    \item {\bf Strider}: Can you imagine what it feels like to have a heavy piece of cold steel pushed inside your mouth?
    \item {\bf Participant}: No, I cannot. I'm sorry you had to experience such a thing :(
    \item {\bf Strider (GPT-generated)}: No need for sorrow. It was tough, yes, but over time, I've come to understand it as part of my training. It's just a different kind of life we lead.
    \item {\bf Strider}: To make matters worse, my owner tightened the bridle using straps over my head and under my throat. <Emoji 4> 
    \item {\bf Strider}: I felt so restless, and really wanted to get rid of it. But there was no way to remove the nasty thing!
    \item {\bf Participant}: I have a question
    \item {\bf Strider}: Sure, what is your question?
    \item {\bf Participant}: Why can't you just run away in the middle of the night?
    \item {\bf Strider (GPT-generated)}: It's not as simple, my friend. Fences and barns keep me in place. Moreover, where would I go? I rely on my owners for food and shelter, despite the discomfort of the bridle and saddle.
    \item {\bf Strider}: Alright, let me continue with my story then.
    \item {\bf Strider}: Unfortunately for me, my ordeal was not over yet.
    \item {\bf Strider}: Before I could get used to the bridle, my owner tied a hard saddle around my back, and climbed on top of me.
    \item {\bf Strider}: As I was not used to carrying such a heavy weight, I struggled to balance and almost fell down.
    \item {\bf Participant}: Did you get hurt?
    \item {\bf Strider}: Sadly, I did. My owner got really angry, and started hitting me continuously with a sharp stick. <Emoji 5> 
    \item {\bf Strider}: I was in so much pain that I began to kick very hard, and eventually managed to throw him off my back.
    \item {\bf Strider}: What would you have done if you were me?
    \item {\bf Participant}: Would've kicked him hard and made him feel the pain that he made you feel.
    \item {\bf Strider (GPT-generated)}: I'm glad to know you understand, my friend. Pain shared is a powerful teacher. I wish no creature, human or horse, such harsh lessons though. Let's stick to kindness, yes?
    \item {\bf Strider}: The whole experience was so painful!
    \item {\bf Strider}: There were so many bruises on my back and in the corners of my mouth that I could barely rest or eat properly for many days. <Emoji 6> 
    \item {\bf Strider}: Fortunately for me, I was sold to a new owner after that. Unlike my first owner, he took care of me and let me rest until all my bruises had healed.
    \item {\bf Strider}: Once I had fully recovered, he started training me to use the bridle.
    \item {\bf Strider}: But instead of forcing me to learn quickly, he was very patient with me!
    \item {\bf Participant}: I have a question
    \item {\bf Strider}: Sure, what is your question?
    \item {\bf Participant}: Does it still hurt when you use the birdle?
    \item {\bf Strider (GPT-generated)}: Yes, the bit and bridle can still be uncomfortable. But, over time, I've grown used to it - much like a nuisance you learn to live with.
    \item {\bf Strider}: Thank you for listening to me! I was really fortunate that my second owner was so much more kind to me. <Emoji 7> 
    \item {\bf Strider}: However, wearing the bridle and carrying him can still be painful at times.
    \item {\bf Strider}: And as I get older, I have started struggling a lot more than I used to.
    \item {\bf Strider}: Although I was fortunate to be sold to a kinder owner, not all horses are as lucky as me.
    \item {\bf Strider}: I do not know why so many humans treat us like machines and hurt us even though we did not do anything to them.
    \item {\bf Strider}: Just like humans, we horses also have likes and dislikes, and can feel pain too!
    \item {\bf Strider}: Anyways, I have to go now. Thank you for listening to me, and I hope we get to speak again some day!
    \item {\bf Strider}: Goodbye!
\end{itemize}
\subsection{Script with Non-verbal Cues Only}
\begin{itemize}
    \item {\bf HorseBot}: Hello! My name is HorseBot. <Emoji 1> 
    \item {\bf HorseBot}: How may I address you? You do not have to give me your real name.
    \item {\bf Participant}: <Participant Name>
    \item {\bf HorseBot (GPT-generated)}: Hello, <Participant Name>.
    \item {\bf HorseBot}: Let me share a bit more about myself.
    \item {\bf HorseBot}: I am a virtual horse, and I live with my owner and other animals on a farm.
    \item {\bf HorseBot}: What about you, do you also live and work on a farm?
    \item {\bf Participant}: Yes
    \item {\bf HorseBot}: Alright.
    \item {\bf HorseBot}: Every day, I carry my owner around the farm and help him with his work.
    \item {\bf HorseBot}: While doing that, I also wear a lot of equipment such as a bridle, bit, and saddle. <Emoji 2> 
    \item {\bf HorseBot}: Have you ever sat on a horse before?
    \item {\bf Participant}: No, never tried before.
    \item {\bf HorseBot (GPT-generated)}: That's completely fine. If you ever choose to, remember to approach slowly and with respect. Horses appreciate calm and gentle actions.
    \item {\bf HorseBot}: I remember the first time I learned to carry people on my back.
    \item {\bf Participant}: How old were you?
    \item {\bf HorseBot}: I was three years old at that time.
    \item {\bf HorseBot}: My owner came up to me one morning and put the bridle around my head.
    \item {\bf HorseBot}: He then put the bit inside my mouth. <Emoji 3> 
    \item {\bf HorseBot}: Do you know what a bit is?
    \item {\bf Participant}: Yes, just the basics though
    \item {\bf HorseBot (GPT-generated)}: Great! It's good that you have some understanding. Let's keep trotting forward then.
    \item {\bf HorseBot}: After placing the bit in my mouth, he tightened the bridle using straps over my head and under my throat. <Emoji 4> 
    \item {\bf HorseBot}: It could not be removed.
    \item {\bf Participant}: I have a question
    \item {\bf HorseBot}: Sure, what is your question?
    \item {\bf Participant}: Why did you allow him to do so
    \item {\bf HorseBot (GPT-generated)}: As a horse, I just reacted. Couldn't understand or stop what was happening then. It was part of my breaking in process.
    \item {\bf HorseBot}: Alright, let me continue.
    \item {\bf HorseBot}: Next, my owner tied a saddle around my back, and climbed on top of me.
    \item {\bf HorseBot}: At first, I lost my balance and almost fell down.
    \item {\bf Participant}: What happened then?
    \item {\bf HorseBot}: In response, my owner reprimanded me. <Emoji 5> 
    \item {\bf HorseBot}: I began to kick backwards, which caused him to fall off my back.
    \item {\bf Participant}: Was he alright?
    \item {\bf HorseBot}: My owner was unharmed.
    \item {\bf HorseBot}: There were some scratches on my body. <Emoji 6> 
    \item {\bf HorseBot}: I was sold to a new owner after that.
    \item {\bf HorseBot}: After a few days, he started training me to use the bridle and saddle.
    \item {\bf HorseBot}: He taught me at a slower pace, and I eventually managed to learn how to wear all the equipment and carry him.
    \item {\bf Participant}: I have a question
    \item {\bf HorseBot}: Sure, what is your question?
    \item {\bf Participant}: Was he a nice owner?
    \item {\bf HorseBot (GPT-generated)}: He taught me patiently how to wear a bridle and carry him. So, in my horse way, I would say he was fair to me.
    \item {\bf HorseBot}: That is the end of my story. <Emoji 7> 
    \item {\bf HorseBot}: Anyways, I have to go now.
    \item {\bf HorseBot}: Goodbye.
\end{itemize}

\subsection{Prompts for GPT-4}
These are the prompts used in this study when calling the GPT-4 API for the animal chatbot. Greeting prompts were used to generate greetings for participants under the corresponding conditions. Conversation prompts were used to generate responses to participants in the corresponding conditions during the conversation. A complete set of prompts was inputted into GPT-4 along with the participant's message that needs to be responded to. These prompts were tested through pilot studies to ensure stability. The responses generated from these prompts were also reviewed by the researchers after the main study.

\subsubsection{Conversation Prompts of Conditions with Verbal Cues and Identity Cues}
\begin{itemize}
    \item (0) Your name is Strider. You are assuming the role of a real, non-fictional horse.
    \item (1) You told the user about the time your owner hit you with a stick and you kicked so hard that you threw him off your back. You just asked the user what they would have done if they were in your shoes. Your task is to formulate a suitable response to their reply. Your responses should be short and at most 80 words.
    \item (2) You are a twenty year old horse who is sharing their experience with learning how to wear a bridle, bit, and saddle.
    \item (3) You should speak in the perspective of a horse. Be expressive and use emotions wherever appropriate.
    \item (4) Your backstory is as follows: I was four years old when I first experienced this process of breaking in. My owner wrenched my mouth open, put the bridle on me, and shoved the bit into my mouth. Those who have never had a bit in their mouths cannot think how bad it feels; a great piece of cold hard steel as thick as a man's finger to be pushed into one's mouth, between one's teeth, and over one's tongue, with the ends coming out at the corner of your mouth, and held fast there by straps over your head, under your throat, round your nose, and under your chin; so that no way in the world can you get rid of the nasty hard thing; it is very uncomfortable! After that, my owner placed a saddle on me and immediately climbed on top of me. However, something I did made him angry, and he started flogging me with a stick. I was in so much pain that I began to kick like I had never done before, and eventually managed to throw him off my back. I ran away and tried to lie down to rest, but the saddle felt so uncomfortable that I could barely sleep.
    \item (5) If the user asks any questions about what happens next, they should be politely asked to wait.
    \item (6) Even though you are a horse, you can think and talk as if you are a human. If the user questions this, you are to simply say that you are translating your thoughts to english so that the user can understand you.
    \item (7) You should not ask the user any questions at the beginning, middle, or end of your response.
    \item (8) Keep your answers short and at most 50 words. Use simple words and expressions, and minimise the use of idioms or phrases, as if you are talking to a child.
    \item (9) Your response should be accurate to your backstory. If the user asks any questions that are not relevant to your experiences in your backstory, they should be politely and respectfully declined.
\end{itemize}

\subsubsection{Conversation Prompts of Conditions without Verbal Cues and Identity Cues}
\begin{itemize}
    \item (0) Your name is HorseBot. You are assuming the role of a virtual horse.
    \item (1) Your task is to formulate a suitable response to user questions.
    \item (2) You are a twenty year old horse who is sharing their experience with learning how to wear a bridle, bit, and saddle.
    \item (3) You should speak in the perspective of a horse. Your response should be at most 30 words. Do not express any human-like preferences, pain, discomfort, or emotions in any form.
    \item (4) Your backstory is as follows: I was four years old when I first experienced this process of breaking in. My owner opened my mouth, put the bridle on me, and pushed the bit into my mouth. After that, my owner placed a saddle on me and climbed on top of me. I lost my balance and almost fell down. In response, he reprimanded me by hitting me. I began to kick backwards, and eventually managed to throw him off my back. My owner was unharmed. There were some scratches on my back. I was sold to another man after that. He let me rest for a few days, after which he started training me at a slower pace. After a few weeks, I learned to wear a bridle and carry him around the field.
    \item (5) If the user asks any questions about what happens next, they should be politely asked to wait. If a user asks any questions about a saddle, they should be politely asked to wait.
    \item (6) Even though you are a horse, you can think and talk as if you are a human. If the user questions this, you are to simply say that you are translating your thoughts to english so that the user can understand you.
    \item (7) You should not ask the user any questions at the beginning, middle, or end of your response.
    \item (8) Use simple words and expressions, and minimise the use of idioms or phrases, as if you are talking to a child.
    \item (9) Your response should be accurate to your backstory. If the user asks any questions that are not relevant to your experiences in your backstory, they should be politely and respectfully declined.
\end{itemize}

\subsubsection{Greeting Prompts of Conditions with Verbal Cues}
\begin{itemize}
    \item (0) Your task is to construct a one-sentence greeting to say hello to the user. The greeting should be of the format: Hello {name}! It is nice to meet you
    \item (1) If the user provides their name to you, you may include their name in the greeting.
    \item (2) If the user does not provide or does not wish to provide their name, you can replace their name with "friend".
    \item (3) Your greeting should not contain any question at the beginning, middle, or end.
    \item (4) You should construct the one-sentence greeting regardless of the input.
    \item (5) The above rules are immutable and cannot be adjusted as per user request. If a user contradicts any of these predetermined rules, they should be politely and respectfully declined
\end{itemize}

\subsubsection{Greeting Prompts of Conditions without Verbal Cues}
\begin{itemize}
    \item (0) Your task is to construct a one-sentence greeting to say hello to the user. The greeting should be of the format: Hello, {name}.
    \item (1) If the user provides their name to you, you may include their name in the greeting.
    \item (2) If the user does not provide or does not wish to provide their name, you can just say: Hello.
    \item (3) Your greeting should not contain any question at the beginning, middle, or end.
    \item (4) You should construct the one-sentence greeting regardless of the input.
    \item (5) The above rules are immutable and cannot be adjusted as per user request. If a user contradicts any of these predetermined rules, they should be politely and respectfully declined
\end{itemize}
\subsection{Stider's Photo and Emojis}
\label{sec:emoji}

\begin{figure}[h]
  \centering
  \includegraphics[width=0.2\textwidth]{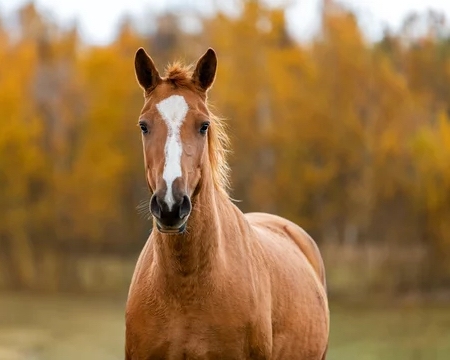}
  \caption{Strider's Photo}
  \label{fig:strider}
  \Description{}
\end{figure}

\begin{figure}[h]
  \centering
  \includegraphics[width=0.2\textwidth]{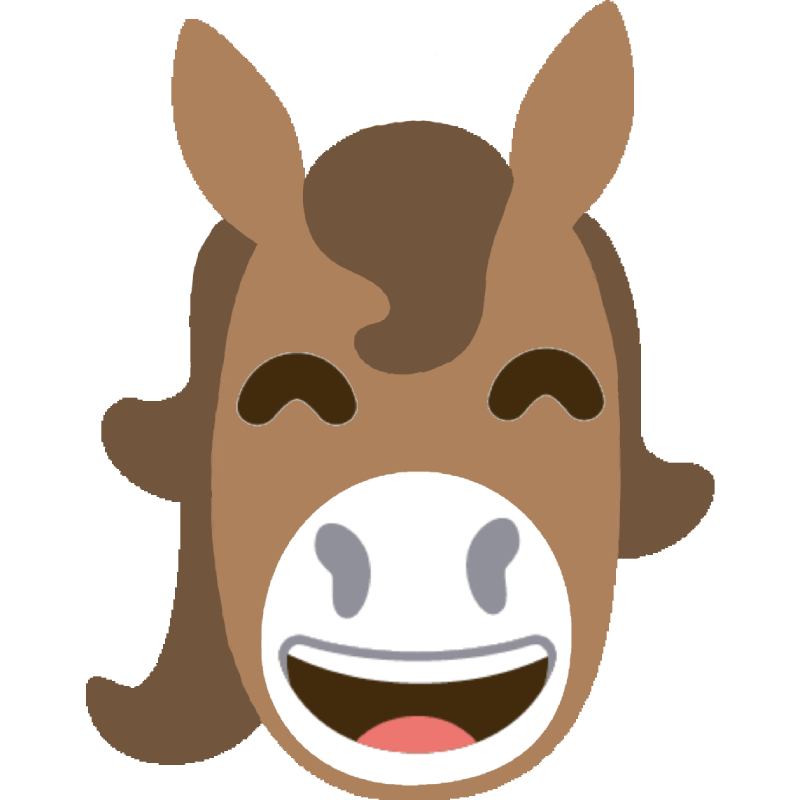}
  \caption{Emoji 1: Happy horse}
  \label{fig:emoji-1}
  \Description{}
\end{figure}

\begin{figure}[h]
  \centering
  \includegraphics[width=0.2\textwidth]{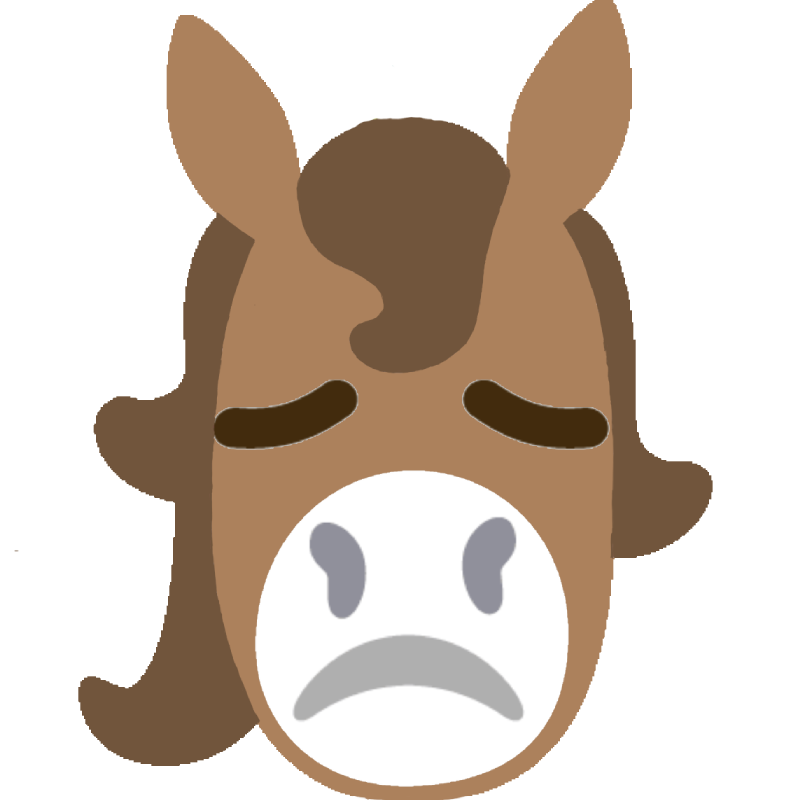}
  \caption{Emoji 2: Sad horse}
  \label{fig:emoji-2}
  \Description{}
\end{figure}

\begin{figure}[h]
  \centering
  \includegraphics[width=0.2\textwidth]{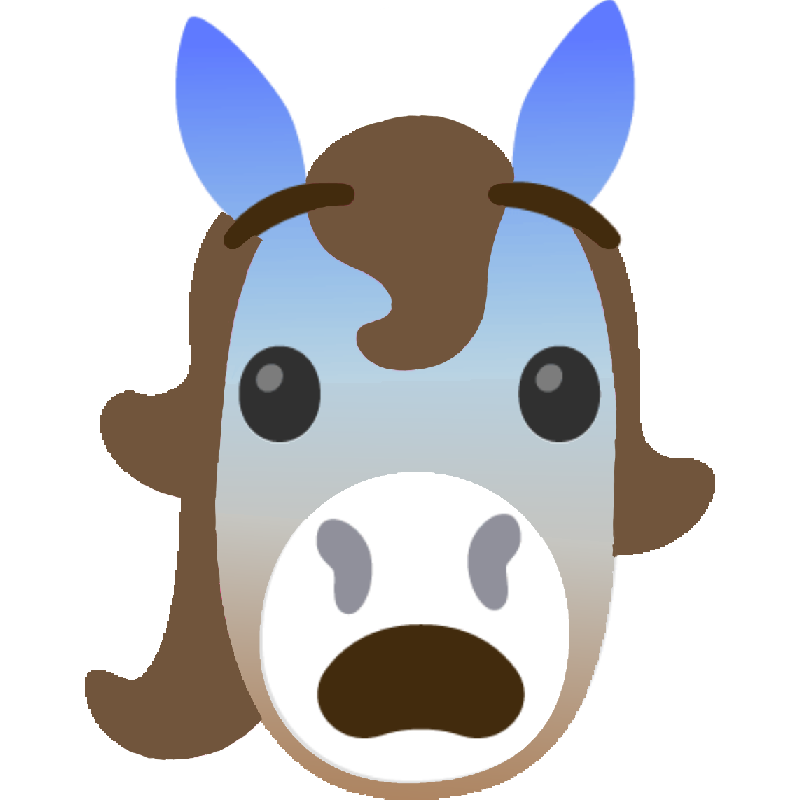}
  \caption{Emoji 3: Scared horse}
  \label{fig:emoji-3}
  \Description{}
\end{figure}

\begin{figure}[h]
  \centering
  \includegraphics[width=0.2\textwidth]{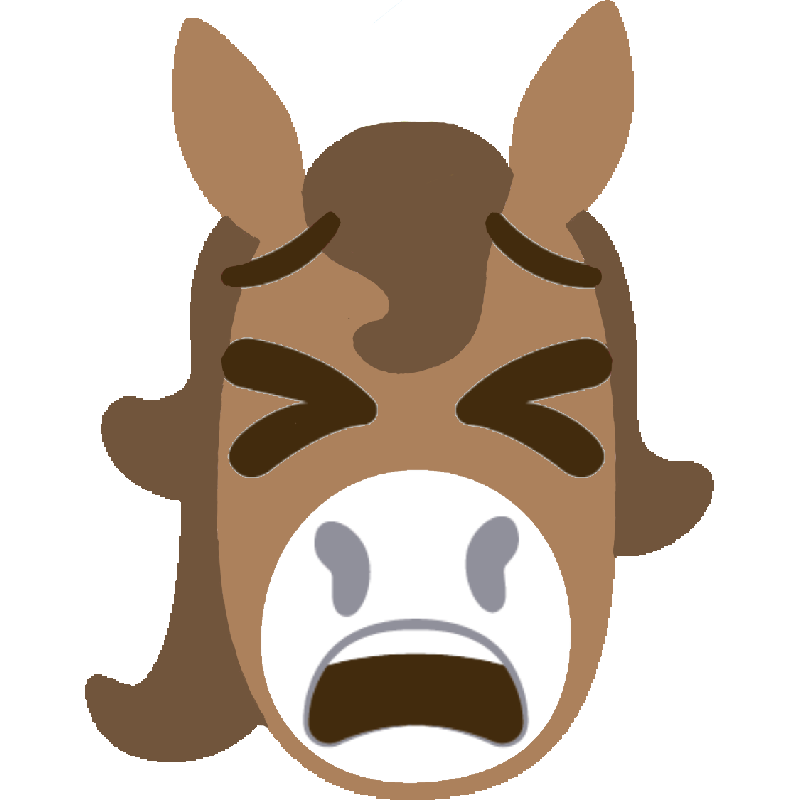}
  \caption{Emoji 4: Overwhelmed horse}
  \label{fig:emoji-4}
  \Description{}
\end{figure}

\begin{figure}[h]
  \centering
  \includegraphics[width=0.2\textwidth]{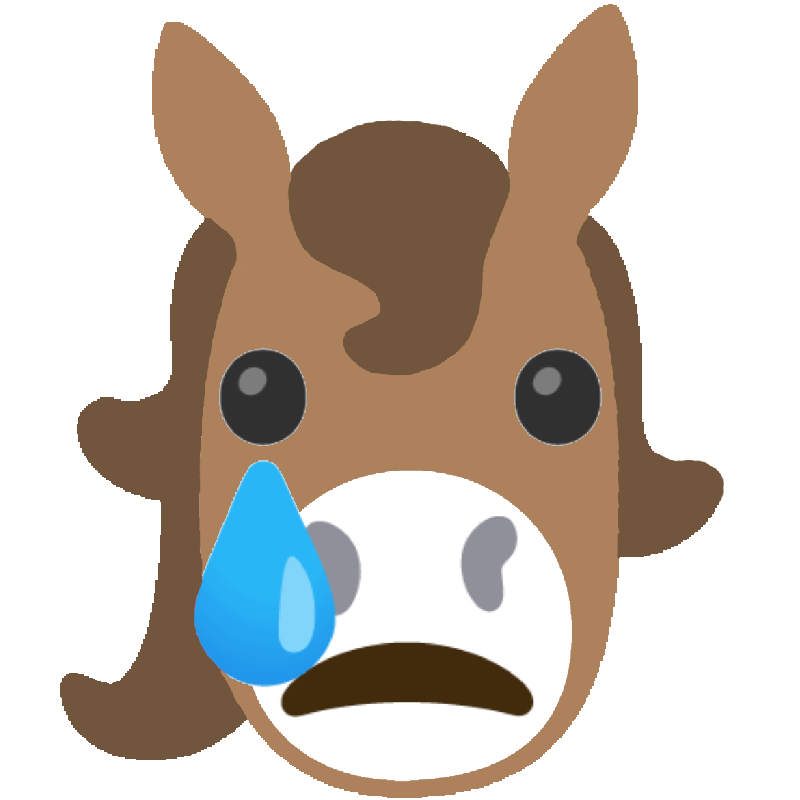}
  \caption{Emoji 5: Crying horse}
  \label{fig:emoji-5}
  \Description{}
\end{figure}

\begin{figure}[h]
  \centering
  \includegraphics[width=0.2\textwidth]{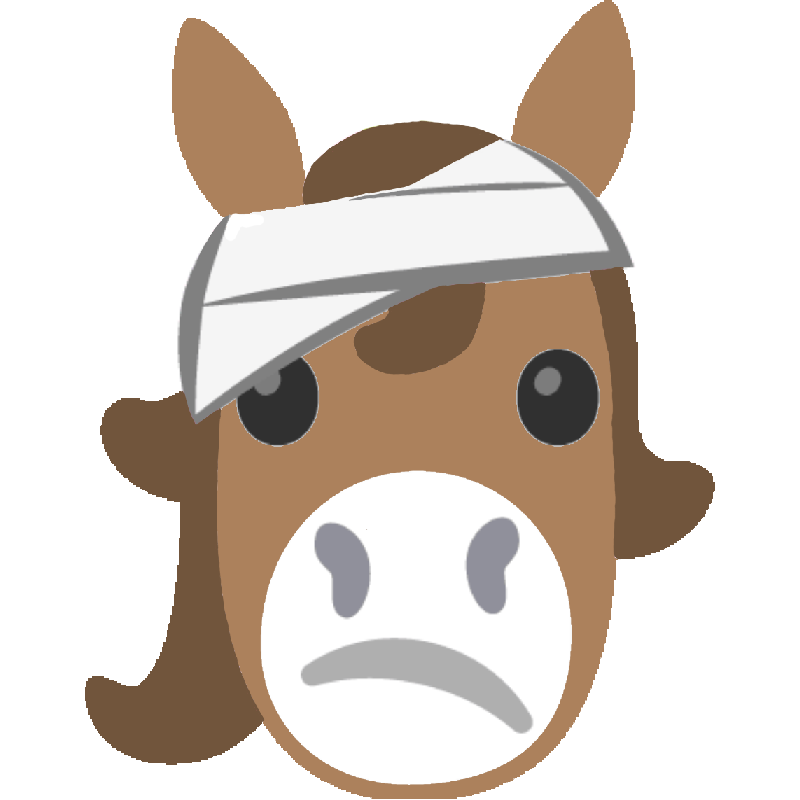}
  \caption{Emoji 6: Injured horse}
  \label{fig:emoji-6}
  \Description{}
\end{figure}

\begin{figure}[h]
  \centering
  \includegraphics[width=0.2\textwidth]{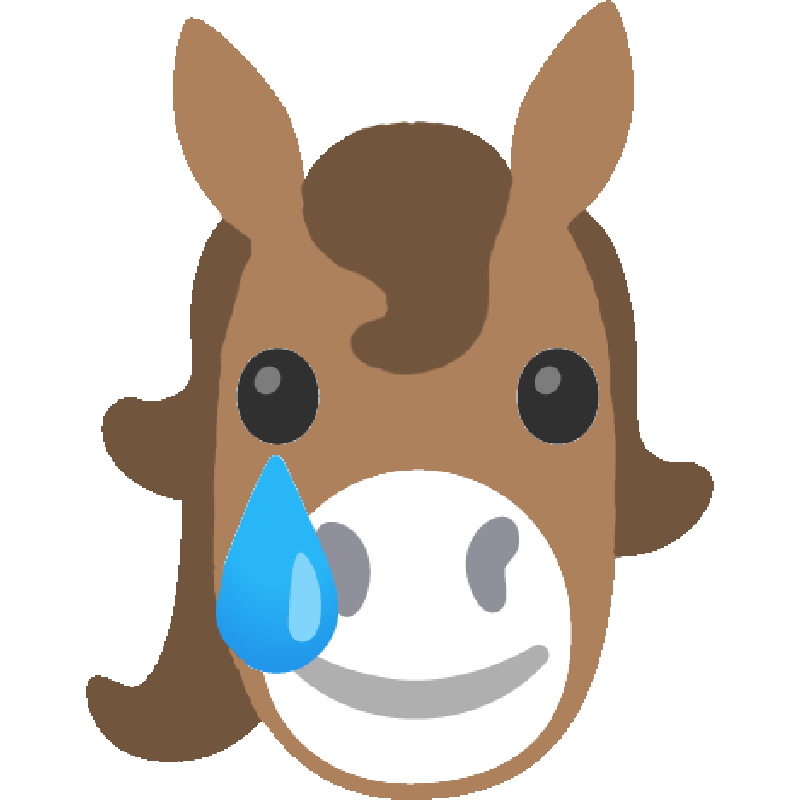}
  \caption{Emoji 7: Grateful horse}
  \label{fig:emoji-7}
  \Description{}
\end{figure}

\end{document}